\documentclass[useAMS,usenatbib]{mn2e}

\usepackage[]{bm}
\usepackage{graphicx}
\usepackage{hyperref}
\usepackage{amssymb}
 
\title[Density Profiles and Double Image Gravitational
  Lenses]{Cusped Mass Density Profiles and Magnification Ratios of Double
  Image Gravitational Lenses}

\author[P. T. Mutka]{P. T. Mutka\thanks{E-mail:Petri.Mutka@oulu.fi}\\
University of Oulu, Department of Physical Sciences, Theoretical
Physics Division, P.O. Box 3000, 90014 Oulun Yliopisto, Finland}

\begin{document}

\date{Submitted 11 March 2009 for
publication in MNRAS by the Royal Astronomical Society and Blackwell Publishing.\\
First revision 10 July 2009.}

\pagerange{\pageref{firstpage}--\pageref{lastpage}} \pubyear{2009}

\maketitle

\label{firstpage}

\begin{abstract}
We have been able to connect the statistics of the observed
double image gravitational lenses to the general properties of the
internal structure of dark matter haloes. 
Our analytical theory for the GNFW lenses with parametrized cusp
slope ($\alpha$) gives us a relation connecting the cusp slope of the lensing
profile to the observed magnification ratio of the produced
images and location of the optical axis. The relation 
does not depend on cosmology, total lens mass,
concentration or redshifts of the the lens and the lensed object.
Simple geometry of axially symmetric lensing and aforementioned
relation enables us to define a threshold value $\alpha_{\mbox{\tiny CSL}}$ for
the cusp slope, independent from location of the optical axis.
The threshold
cusp slope value  $\alpha=\alpha_{\mbox{\tiny CSL}}$ is the shallowest slope for
the inner part of the GNFW profile that can produce the
observed magnification ratio with any lensing configuration.
We use distribution of these threshold values in
a statistical study of the double
image lenses in order to limit the possible cusp slope
values, and identify whether there exists a population of haloes with
similar profiles.
Our theoretical fit indicates that within our sample of double image
gravitational lenses, most of the haloes have cusp slope
$\alpha=-1.95\pm 0.02$. We have also found an indication of
a second population of lenses with 
a cusp slope value $\alpha= -1.49\pm 0.09$. We estimate that there
is about 99 per cent probability that the observed feature in the
threshold value limit distribution is produced by the 
second population of lenses, with their own characteristic density profile.
The data indicating the exact
characteristics of the sub-population is noisy. Roughly one
out of six haloes within the sample belong to this shallower cusp
slope group. We investigate errors
in our analysis by constructing mock catalogues with the
Monte-Carlo method.
\end{abstract}

\section{Introduction}

The large scale structure formation in the universe is dominated
by the gravitational evolution
of dark matter (DM) and the expansion of the universe
\citep{peeb1982,blum1984,davi1985}. Currently,
the best candidate for the DM component is a weakly interacting
massive particle (WIMP).
A host of candidate particles can be suggested from the various
extensions of the standard particle theory, for example
supersymmetry \citep{elli1991,mart1998} or Kaluza--Klein
concepts \citep{chen2002,serv2003}.

Theories concerning structure formation usually postulate
non-interacting dark matter that is affected only by gravity.
Since the last decade, a large number of studies
employing analytical theories and/or N-body simulations have been conducted
on this basis. There are also several variants of this approach, in which
DM has been attributed with additional properties, such as being
self-interacting \citep[e.g.][and references therein]
{sper2000,burk2000,rome2001,dong2003a,dong2003b},
decaying \citep[e.g.][]{adbe2008,borz2008}
or having a postulated equation of state \citep[for example][]{aust2005}.

At the linear regime, the current state-of-the-art cosmological theory,
$\Lambda CDM$-model with inflation, is very successful
and can explain most of the observations. However, problems arise
at the nonlinear regime, where the DM halo is decoupled from the
expansion of the universe and evolves through self-gravitation and
gravitational interactions with its environment
\citep{moor1998,moor1999,powe2003}.

At the high density region of dark matter halo cores, the
physical properties of the dark matter particles and their interactions
should become important. According to  \cite{sper2000,burk2000}
the weak-interaction cross section of the dark
matter particle can have significant effect on the shape of the
density profile on cosmological time scales. The cross section
has also been constrained by \cite{rand2008} using
the Bullet galaxy cluster.
On the other hand, \cite{beac2007} determine upper
limit for the weak-interaction cross section from the cosmic
diffuse neutrino background, and argue that dark matter haloes
cannot be significantly altered by dark matter particle
annihilations. It is clear, that observations and
theoretical studies of dark matter
haloes can help us narrow down the
properties of the candidate particles.

The study of dark matter halo formation
at the nonlinear regime by \cite{nava1996,nava1997} found
that the dark matter haloes follow roughly a
universal radially symmetric density profile (NFW). In the
study, one halo consisted of several thousand gravitating
DM-particles. Similar N-body experiments have been conducted
regularly with increasing number of DM-particles and better resolution by
several authors, see e.g. \cite{moor1998,moor1999,ghig2000}. 

Since then, several N-body study inspired
models for the DM-haloes have been proposed.
They all share some common properties; at the outer fringes, the halo follows
$\rho\propto r^{-3}$ profile. After the transition region at scale
radius $r_{\mbox{\tiny s}}$, the profile is changed to value $\rho\propto r^\alpha$
representing a cusped (or flat) core.

An ad hoc explanation for this behaviour is that the outer
region is in the state of 'inflow', where the dark matter
decouples from the general expansion of the universe, and streams
towards the core region of the halo. This is characterized by
dominating radial component of the velocity field.
Inside the scale radius $r_{\mbox{\tiny s}}$, where the velocity field
is more thermalized, the halo is composed of captured 
dark matter component that has passed through the core and turned
back at least once. See for example \cite{dehn2005}
or \cite{hans2009} for further details.

In their study, \cite{nava1996,nava1997} found an universal 
cusp slope $\alpha=-1$, valid for haloes at extensive range of
size scales. More recent N-body studies have
resulted an array of different slope values ranging from
$\alpha=-1$ to $\alpha=-2$. This ambiguity in resolving the
cuspiness of the DM haloes with N-body simulations, and
contradicting observations from galactic dynamics has been
one of the central weaknesses of the $\Lambda$CDM paradigm.

The observed shape of the density profile can be measured indirectly by
analyzing internal dynamics and kinematics of nearby galaxies,
measuring X-ray temperature profile, using Sunyaev-Zel'dovich effect,
or with weak lensing measurements of galaxy clusters.
Several authors has published results characterizing the profile shape by
combining some of these observations, see e.g. \cite{mahd2007} and
references therein.

These studies 
either assume a profile shape in advance or try to acquire a fit for
it. In both cases, the NFW profiles ($\alpha\sim-1.0$) have proven
successful \citep{gava2005,point2005,vikh2006,limo2008},
although
shallower density profiles $\alpha\sim -0.5$ have been reported
\citep{sndr2005,voig2006}. On the other hand, lensing
studies by \cite{sere2009} estimated slightly steeper
density profile for the cluster AC114 than the canonical NFW value.

Rotation curves and analysis of dynamics for late type spiral galaxies
indicate that they must have shallower non-cusped cores
\citep{vale2007,dona2009}.
If a galaxy has a bar, it will interact
with the halo and this should slow-down the bar rotation,
see e.g. \cite{wein1985}.
It has been claimed that bars rotate so fast (corotation resonance close
to bar radius) that they could not be embedded into strongly concentrated halos.
On the other hand, the interaction between the bar and halo have been suggested
to make the cusp more shallow \citep{deba2000}. However, there
seems to be possible issues with the resolution of N-body models
\citep{wein2007},
see also \cite{sell2008}.

Interestingly, in recent high resolution  
N-body models by \cite{dubi2009}, the bar stayed fast and did
not destroy the cusp. 
Further complications are provided by recent results indicating that
some small bars rotate slowly \citep{raut2008,chem2009}.
Slow rotation for such bars was suggested already by \cite{elme1985}
on morphological grounds, so it is possible that their origin is somehow
different from larger bars that are studied more frequently.

The dwarf galaxies at low surface brightness are considered as being
dominated by dark matter component. However, direct measurements
of the rotation curves of these objects indicate a constant density core
without a cusp, see \cite{burk1995} or
\cite{zack2006}, and corresponding references therein.

The newest development in the constraints for the dark matter halo
profiles comes from the observational particle physics.
The WMAP probe has observed a microwave ``haze'' around the Galactic
core. This can be explained by hypothetical
dark matter particles annihilating
at the high density region of the Milky Way core.
The ``haze'' is interpreted as emitted synchrotron
radiation from the produced electron-positron pairs, and it
matches predicted emission from a dark matter halo with a 
cusped profile, $\rho\propto r^{-1.2}$ within the inner
kiloparsecs \citep{hoop2007}. Since there is no
definite direct detection of the dark
matter particle to date, this can be considered as a
hypothetical scenario yet to be proved or disproved.

The largest coherent sample of the galaxy scale
strong gravitational lenses has been measured with the HST
in the Sloan Lens ACS survey \citep{bolt2006}. \cite{gava2007} made a
statistical study about subset of these lenses, and confirmed that on average,
an isothermal profile $\alpha\sim -2$ fits the lensing profiles.
Several strong gravitational lensing studies have also suggested
shallower cusp slopes \citep{sand2002,sand2004}
at $\alpha\sim -0.5$.

However, flat profiles have severe problems in strong lensing.
In our previous paper \cite{muma2006}, we developed a
semi-analytical theory for axially symmetric gravitational lensing
and employed it to the generalization of the NFW profile (GNFW),
by \cite{nava1996,nava1997,zhao1996}. In the GNFW profile, value of
the cusp slope is a free parameter:
\begin{equation}\label{profile}
\rho\propto r^\alpha(1+r)^{-3-\alpha}.
\end{equation}
Note that we use convention in which the cusp slope $\alpha$
has negative values.
In axial lensing context, the GNFW halos with cusp slope
shallower than  $\alpha\sim -1$ have severe problems in strong lensing
\citep{muma2006}.

  
\begin{figure}
\center\includegraphics[width=1.0\linewidth]{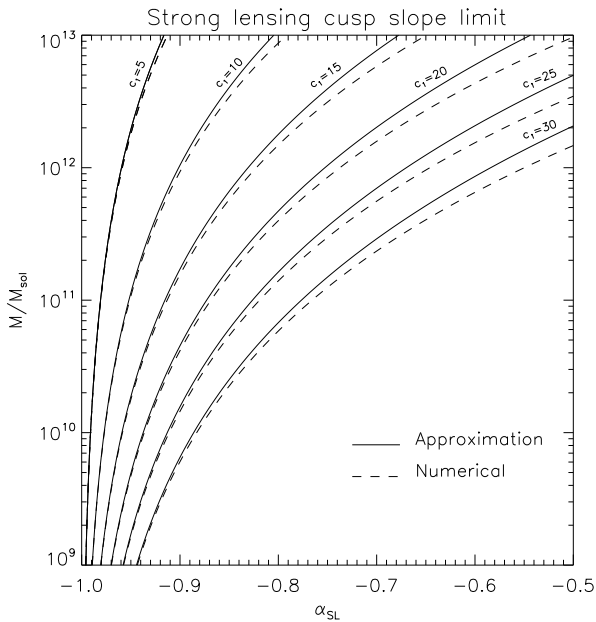}
\caption{\label{cusplimfig}
The strong lensing condition, 
equation (\ref{cusplim}) is function
of the mass $M$ of the lens. This plot shows the required mass
for the lens as a function of the
strong lensing cusp slope limit $\alpha_{\mbox{\tiny SL}}$.
The strong lensing is not possible on the
right hand side (below) of each curve. The curves with 
concentration parameter values $c_1=5,10,15,20,25,30$
are plotted for the lens at redshift $z=0.3$, and the source at
redshift $z=1.0$.  
The dashed line is a numerical version of the condition
approximated by equation (\ref{cusplim}) as prescribed in
the appendix \ref{cusplimderiv}. 
}
\end{figure}
  

The cusp slope limit for strong lensing $\alpha_{\mbox{\tiny SL}}$
is a function of the
cosmological parameters, lensing geometry, lens mass and lens
concentration. Basically, at this limit the radius of the
Einstein ring for a GNFW
lens goes to zero, after which the strong lensing
configurations become impossible.
A good approximation for this limit is
\begin{equation}\label{cusplim}
\alpha_{\mbox{\tiny SL}}=-\frac{3}{2}+\frac 12\sqrt{
1+\frac{1600 r_{\mbox{\tiny s}}\rho_{\mbox{\tiny cr}}c_1^3(1+c_1)}{3\Sigma_{\mbox{\tiny cr}}[(1+c_1)\log(1+c_1)+c_1]}
},
\end{equation}
that produces values slightly above the $\alpha\sim -1$, with
reasonable lens masses and concentrations. Figure
\ref{cusplimfig} shows the required lens mass as function of the
strong lensing condition, calculated with equation (\ref{cusplim}) and
corresponding fully numerical solution. With usual lens masses order
of $M\sim 10^{12} M_{\odot}$, the strong lensing is limited close to
$\alpha\sim -1$. Only excessive concentration allows lens systems with slope
$\alpha\sim -0.5$ that can produce multiple images.
Appendix \ref{cusplimderiv} presents definitions of $r_{\mbox{\tiny s}}$,
$\rho_{\mbox{\tiny cr}}$ and $\Sigma_{\mbox{\tiny cr}}$ and derivation for the equation (\ref{cusplim}).

It can also be shown that, if dark matter haloes have
shallower profile ($\alpha\ge -1.5$), triple
image lenses with visible inner caustic image
should be very common. The estimated probability
for a triple image lens configuration is
\begin{equation}\label{pimpom}
P_{\mbox{\tiny triple}}(\alpha)=1-\frac{\alpha^2}{4},
\end{equation} 
and there are only three observed lens systems with (suspected)
visible inner caustic image out of one hundred.
This gives an average cusp slope value
$\alpha\sim -1.97$, which is closer to the isothermal value.
See section \ref{tripleimage} for derivation of the equation
(\ref{pimpom}) and further details.
Note that equations (\ref{cusplim}) and (\ref{pimpom}) are also valid with
reasonable values of eccentricity for elliptic lens haloes.
See section \ref{theory} for further discussion.

We investigate the characteristics of the lensing profiles by deriving
a  cusp slope limit (CSL) value $\alpha_{\mbox{\tiny CSL}}$ for the shallowest
cusp slope, that can produce the required flux difference for the images
with any lensing geometry (i.e. it must hold
$\alpha < \alpha_{\mbox{\tiny CSL}}$ for the lensing profile).
We assume random alignments for the
sources, and use the CSL
value in statistical analysis of the double image lenses.
By examining distribution of
these threshold values, we can characterize
the general properties of lensing profiles.
For this purpose, we composed a catalogue of
the double image lenses exhibiting
properties of radial symmetry. Our sample is based on the
known cases of the lensed quasars listed in CASTLES survey.

Our method relies on the observed flux ratios of double image
lenses. Because we use image flux ratios, information on the lens and
source redshifts, lens mass and concentration, and cosmology is not
required. With our formulation, the intrinsic halo properties are
separated from these quantities. Therefore, the statistical method
does not require theoretical distributions characterizing lens and
source populations (such as Schechter function coupled
with the mass-luminosity
relation for lenses, or quasar luminosity functions for sources),
which makes it possible to avoid number of
different uncertainties in the analysis.

However,
there are several factors that can corrupt the observed
fluxes from the predicted magnification produced
by the lens. We examine effects by the lens ellipticities,
the substructure in the lensing halo and
time-delay from a variable source.
This is done by constructing mock lens catalogues with
the Monte Carlo method when accounting for these factors.

The next section presents a brief review of our lensing theory,
formulation for the CSL value, triple image lens probability and
description for theoretical CSL distribution. The third section
describes the selected lens systems and the catalogue used in the study.
The Monte Carlo testing of the 
CSL statistics is presented in section four.
Finally, we employ the statistical CSL analysis on real double image
lens data in section five. The last section contains conclusions
and discussion regarding the study. The derivation
for the cusp slope limit for strong lensing
is presented in appendix \ref{cusplimderiv}.

\section{Theory}\label{theory}

  
\begin{figure}
\center\includegraphics[width=1.0\linewidth]{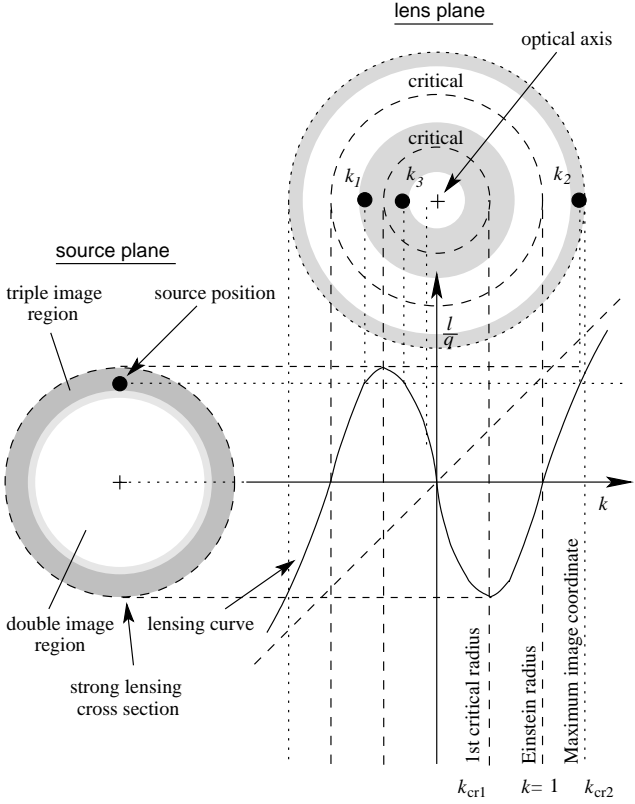}
\caption{\label{f1eps} The lens equation normalized with the Einstein radius
  (\ref{leq}) produces three images $k_1$, $k_2$ and $k_3$. The first
  image $k_1$ is located between the Einstein radius ($l/q=0$ or
  $k=1$) and the first
  critical radius $k=\nu$. The second image is at the opposite
  side of the optical axis, between the Einstein radius and the second
  critical radius, equation (\ref{crirad2})
  (the largest possible distance from the optical axis
  for the multiple imaged source). The third image $k_3$
  (inner image) is inside the first critical radius, and it is usually
  strongly demagnified. With large values of $k$, outside the second
  critical radius, the lensing curve should approach to $l/q=k$. The
  chosen lensing curve strongly exaggerates size of the triple image
  region.
}
\end{figure}
  

Here we briefly review the relevant lensing theory presented in our
previous paper \citep{muma2006}, and derive the cusp slope limit (CSL)
value. The CSL value is based on the analytical formulation of the
magnification ratio of the lensed images, and its solutions.

In gravitational lensing, the source image is lensed by the
gravitational field of the lens object. The source image that is
thought to be confined on a plane at the source distance
$D_{\mbox{\tiny S}}$, produces
observed images on the lens plane (at the lens object
distance $D_{\mbox{\tiny L}}$).
Mapping from the source plane to the lens plane is described
by the lens equation.

In axially symmetric lensing, we choose the line connecting the
observer and the lens as origin for the frame of
reference. Corresponding radial coordinate at the source plane is
denoted with $y$ and at the lens plane with $x$, as in \cite{schn1992}.

When the source is located at the origin ($y=0$), it is mapped to the
lens plane as an
Einstein ring, that has radius $x=x_e$.
The axially symmetric lens equation for the GNFW density profile,
normalized with the Einstein radius $x_{\mbox{\tiny e}}$, can be written as
\begin{equation}\label{leq}
l=\left\{\begin{array}{llll}
qk\left(1-|k|^{\alpha+1}\right)
&\mbox{if}\,\alpha\ne -1, -2\leq\alpha<0 \\
\frac{\mu_0}{2}k\log|k| &\mbox{if}\,\alpha=-1
\end{array}\right.,
\end{equation}
where the radial coordinate at the source plane is
$l=y/(x_{\mbox{\tiny e}}r_{\mbox{\tiny s}})$ and corresponding radial
coordinate at the lens plane is 
$k=x/x_{\mbox{\tiny e}}=r/(r_{\mbox{\tiny s}}x_{\mbox{\tiny e}})$.
Scale radius $r_{\mbox{\tiny s}}$, and constants $q$ and
$\mu_0$ depend on the cosmology, the mass and the concentration
of the lens object, and
the geometrical setup of the lensing. Note that our cusp slope value
is a negative number. The exact definition for the scale
  radius $r_{\mbox{\tiny s}}$ and
the corresponding unnormalized lens equation can be found in appendix
\ref{cusplimderiv} and paper \cite{muma2006}.

With this notation, the
Einstein radius is located at $k=1$. The first
critical radius at $k_{\mbox{\tiny cr1}}=\nu=(\alpha+2)^{-1/(\alpha+1)}$
and the corresponding twin image at the opposite side of the
lens has a good approximation
\begin{equation}\label{crirad2}
k_{\mbox{\tiny cr2}}=\frac{\alpha+1+\sqrt{1+2\nu}}{\alpha+2}.
\end{equation}
The critical radii ($k=1$ and $k=k_{\mbox{\tiny cr1}}=\nu$)
are connected to the caustic curves where the
magnification diverges, see figure \ref{f1eps}, or \cite{muma2006}
for further details. Note that although the image at
$k_{\tiny\mbox{cr1}}$ is at the critical curve, the corresponding
twin image at $k_{\tiny\mbox{cr2}}$ is not.

The approximation in the lens equation (\ref{leq})
breaks down with large values of the radial coordinate $k$, and therefore
the lens equation (\ref{leq})
does not behave asymptotically as it should.
This asymptotic behaviour can be corrected by writing the lens
equation ($\alpha\ne -1$) piece-wisely as
\begin{equation}\label{leq2}
l=\left\{\begin{array}{lll}
qk(1-|k|^{\alpha+1}), &\mbox{when}\;\; k\le k_{\mbox{\tiny B}} \\
k-\frac{\alpha+1}{\alpha+3}\frac{k_{\mbox{\tiny B}}^2}{k}(1-q),\; &\mbox{when}\;\; k>k_{\mbox{\tiny B}}
\end{array}\right.,
\end{equation}
where the lens equation (\ref{leq2}) is divided at
\begin{equation}
k_{\mbox{\tiny B}}=\left(\frac{2(q-1)}{q(\alpha+3)}\right)^{1/(\alpha+1)}
\end{equation}
in order to avoid negative surface densities.

A lensed image at location $k\leq k_{\tiny\mbox{B}}$ is magnified
  by the lens as 
\begin{equation}
\mu=
q^{-2}\left|1-|k|^{\alpha+1}\right|^{-1}
\left|1-(\alpha+2)|k|^{\alpha+1}\right|^{-1},
 \ \alpha\ne-1.
\end{equation}
Note, that the factor $q$ dependency is contained into a coefficient
multiplying the magnification. Therefore, the $q$ dependency is
cancelled out when considering ratio of image magnifications in strong
lensing cases.

The positive definitive
lensed images at $k_1$, $k_2$ and $k_3$ that are
solutions of the lens equation (\ref{leq}) or (\ref{leq2}), are labelled
as in figure \ref{f1eps}.
The image closest to the source, that
has positive parity, is located at $k_2$.
The images at $k_1$ and $k_3$ are at the opposite
side of the lens with negative and positive parities
(the image at $k_3$ is usually strongly demagnified).
The solutions $k_1$ and $k_2$ of a source at $l$ in the lens equation
must satisfy
\begin{equation}\label{modsol}                                         
\frac{k_1^{\alpha+2}+k_2^{\alpha+2}}{k_1+k_2}=1,                       
 \ \mbox{when}\, \alpha\ne -1.
\end{equation}                                                         
Now we can parameterize the lens equation with coordinate ratio
$\theta=k_1/k_2=x_1/x_2=\theta_1/\theta_2 \leq 1$, where
$\theta_1$ and $\theta_2$ are observed angular distances of the images    
from the optical axis on the sky correspondingly. The advantage of the    
parameter $\theta$ is that it is directly observable, if the location
of the optical axis is known.

The ratio of the image coordinates
$\theta=k_1/k_2$ has a minimum value at
\begin{equation}
\theta=\theta_{\mbox{\tiny min}}=\frac{k_{\mbox{\tiny cr1}}}{k_{\mbox{\tiny cr2}}}=\frac{\nu(\alpha+2)}{\alpha+1+\sqrt{1+2\nu}}
\end{equation}
corresponding to the most asymmetric configuration for the strong
lensing for a source image at $l=l_{\mbox{\tiny max}}$.
If the image coordinate ratio $\theta$
is decreased below this value, it describes the ratio
$\theta=k_3/k_2$ for the demagnified inner caustic image. Thus,
$\theta_{\mbox{\tiny min}}$ is also the smallest value for the $\theta=k_1/k_2$.

When relation (\ref{modsol}) is employed, and $\alpha\ne -1$,
the image coordinates become                                   
\begin{equation}\label{thetak}                               
\begin{array}{llllll}                                        
k_1 & = &                                                    
\theta\left[\frac{1+\theta}{1+\theta^{\alpha+2}}\right]^{1/(\alpha+1)}
\\                                                                    
k_2 & = &                                                             
\left[\frac{1+\theta}{1+\theta^{\alpha+2}}\right]^{1/(\alpha+1)}      
\end{array}                                                          
\end{equation}
and the lens equation can be written as
\begin{equation}\label{thetal1}                                                
l=q\theta\left[\frac{1+\theta}{1+\theta^{\alpha+2}}\right]^{1/(\alpha+1)}      
\frac{\theta^{\alpha+1}-1}{\theta^{\alpha+2}+1},\;\alpha\ne -1, \
k<k_{\tiny\mbox{B}}.                
\end{equation}

Now the magnification ratio $M$ of the lensed images as a
function of the cusp slope $\alpha$
and ratio of the image coordinates $\theta=k_1/k_2$ becomes
$$
M(\alpha,\theta)=\frac{\mu_1}{\mu_2}
$$
\begin{equation}\label{opaxis}
=\theta\frac{
|\theta^{\alpha+2}+1-(1+\theta)(\alpha+2)|
}{
|\theta^{\alpha+2}+1-\theta^{\alpha+1}(1+\theta)(\alpha+2)|
}
, \ \alpha\ne -1, \
k<k_{\tiny\mbox{B}}
\end{equation}
as presented in \cite{muma2006}.
Note that the derived relation (\ref{opaxis})
between the location of the optical
axis, the image magnification ratio and the value
of the cusp slope does not depend on the mass or the
concentration of the lens, the cosmological model, or the redshifts
of the source and lens object. This makes it an ideal tool for
studying the structure of the lensing halos by separating their intrinsic
properties from the extrinsic lensing conditions.

When the magnification (flux) ratio of the observed images ($M_0$) is
measured,  the location of
the optical axis ($\theta$) as a function of the cusp slope parameter
$\alpha$ can be solved from equation (\ref{opaxis}). 
Basically, there are two different groups of solutions
corresponding to triple and double image lens configurations.

  
\begin{figure}
\center\includegraphics[width=1.0\linewidth]{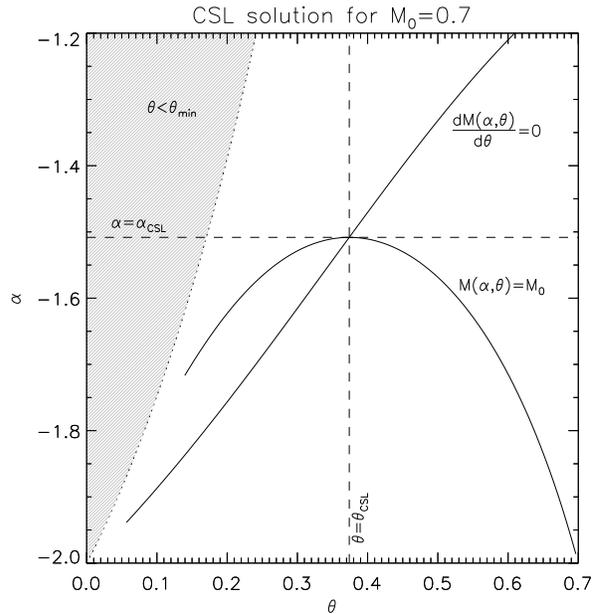}
\caption{\label{f3eps}
The CSL solution for a double image lens system with image flux ratio
$M(\alpha,\theta)=M_0=0.7$, as calculated from equations (\ref{equation1}) and
(\ref{equation2}). Horizontal axis shows coordinate ratio
$\theta=k_1/k_2$ and vertical axis profile slope $\alpha$. Curve
$\mbox{d}M/\mbox{d}\theta=0$ intersects the magnification ratio
solutions $M(\alpha,\theta)=M_0$
at the shallowest possible cusp slope value
$\alpha=\alpha_{\tiny\mbox{CSL}}$.
The shaded area corresponds to the inner image solutions, where
$k_1$ is replaced by $k_3$, and triple image solutions are limited
nearby $\theta=\theta_{\tiny\mbox{min}}$.
}
\end{figure}
  

If the solutions are plotted at ($\alpha,\theta$)-plane,
the double image solutions vanish at $\alpha=\alpha_{\mbox{\tiny CSL}}$ 
and $\theta=\theta_{\mbox{\tiny CSL}}$, where
\begin{equation}\label{equation1}
\frac{\mbox{d}M}{\mbox{d}\theta}_{\Big|_{
\scriptsize
\begin{array}{lll}
\theta=\theta_{\mbox{\tiny CSL}}\\
\alpha=\alpha_{\mbox{\tiny CSL}}
\end{array}
}}=0.
\end{equation}
This means that the double image solutions are restricted to cusp slopes
$\alpha\in[-2,\alpha_{\mbox{\tiny CSL}}]$.
But on the other hand, using the relation (\ref{opaxis}),
this point can be related to the observed magnification ratio
with
\begin{equation}\label{equation2}
M(\alpha_{\mbox{\tiny CSL}},\theta_{\mbox{\tiny CSL}})=M_0.
\end{equation}
Pair of equations (\ref{equation1}) and (\ref{equation2}),
can be used to solve $\theta_{\mbox{\tiny CSL}}$
and $\alpha_{\mbox{\tiny CSL}}$ for a
lens system with observed magnification ratio
$M_0=\mu_1/\mu_2=F_1/F_2$ (where $F_1$ and $F_2$ are corresponding
observed fluxes for the images). See figure \ref{f3eps}.

At the configuration $(\alpha_{\mbox{\tiny CSL}},\theta_{\mbox{\tiny CSL}})$,
the lens has just enough 'power' to produce
the magnification difference for the observed flux ratio.
For shallower cusp slope values $\alpha>\alpha_{\mbox{\tiny CSL}}$
the observed flux ratio cannot be reached with any double image configuration.

The absolute maximum for the
possible values of the cusp slope $\alpha$ can be calculated from the
measured flux ratio for each observed
double image lens system.
This is the previously discussed cusp slope limit, abbreviated as CSL.
When we assume random lensing alignments, the
distribution of the CSL values depend only on the properties of the
lensing profiles. For a single universal cusp slope value, the
numerical formulation for the CSL probability is presented
in section \ref{CSLdistro}.

\subsection{Condition for triple image lensing}
\label{tripleimage}

An axially symmetric lens potential produces visible triple image
lens when images at $k_1$ and $k_3$ are sufficiently close to the
radial critical
curve at $k_{\mbox{\tiny cr1}}=\nu$. Otherwise, the image $k_3$ inside the caustic
curve becomes unobservable because of strong demagnification, while
image $k_1$ remains magnified, and a double image lens system is observed.

Images nearby the critical radius $k_{\mbox{\tiny cr1}}$ are produced by source images
that are located near $l\sim l_{\mbox{\tiny max}}$, where $l_{\mbox{\tiny max}}$ is the maximum
source coordinate for producing multiple images (strong lensing).
Thus, the triple image condition demarcates out part of the strong
lensing region. In other words, for double image lensing, the source
coordinate $l$ must have $l<l_{\mbox{\tiny triple}}<l_{\mbox{\tiny max}}$.

The condition for triple image lensing can be derived by examining the
magnification ratio $\mu_1/\mu_3$ of the images $k_1$ and $k_3$. This
is done by expanding the lens equation at $k=\nu$, setting
the condition $\mu_1/\mu_3<\tau$ and solving it for the corresponding
source coordinate. Here $\tau$ is the limiting value for the observable
magnification ratio.

From the lensing theory presented in \cite{muma2006}, the maximum source
coordinate has value
\begin{equation}\label{maxsource}
L_{\mbox{\tiny max}}=\left(\frac{l}{q}\right)_{\mbox{\tiny max}}
=-\frac{\alpha+1}{\alpha+2}\nu,
\end{equation}
where $\nu=(\alpha+2)^{-1/(\alpha+1)}$ and $\alpha$ is the cusp slope
of the density profile in equation (\ref{profile}).

We normalize the lens equation
\begin{equation}
\frac lq=k(1-|k|^{\alpha+1})
\end{equation}
with expression (\ref{maxsource}) in case of $-2\le \alpha < -1$. 
Now the lens equation can be
expressed in new coordinates $y=l/(qL_{\mbox{\tiny max}})$ and $x=k/\nu$ as
\begin{equation}
y=\frac{x}{\alpha+1}(\alpha+2-x^{\alpha+1}).
\end{equation}
We continue our derivation by expanding the re-normalized lens equation at
$k=k_{\mbox{\tiny cr1}}=\nu \Leftrightarrow x=1$, which gives us
\begin{equation}
y\approx 1-\frac{\alpha+2}{2}(x-1)^2.
\end{equation}
By solving this approximation we get the image location
\begin{equation}\label{solutions}
x=1\pm\sqrt{2\frac{1-y}{\alpha+2}}
\end{equation}
from a source nearby the critical curve.
Here the positive sign corresponds to image $k_1$ and negative to
$k_3$. Again, from the lensing theory presented in \cite{muma2006},
the magnification ratio condition for the triple image lensing can be
expressed as
\begin{equation}
\frac{\mu_1}{\mu_3}=\left(\frac{x_1}{x_3}\right)^2\frac{|y-x_3|}{|y-x_1|}<\tau.
\end{equation}
By inserting solutions for $x_1$ and $x_3$ from equation
(\ref{solutions}) to this expression and solving for the source
coordinate $y$, 
the condition for observable triple image lensing becomes
\begin{equation}
y>1-\frac{\alpha+2}{2}\left(\frac{\tau-1}{\tau+1}\right)^2.
\end{equation}
The power of two factor in parenthesis that depends on $\tau$,
approaches very quickly to the unit value if demanding ten or a hundredfold
magnification ratio for the limit of observability. Because
determining the exact value for the magnification ratio limit is
impossible, we set this factor equal to one. Thus,
practically without any loss of accuracy, the source image condition for
the triple image lensing can be generalized to
\begin{equation}\label{triplecond}
y>-\frac{\alpha}{2} \Leftrightarrow
\frac{l}{q}>\frac{\alpha\nu}{2}\frac{\alpha+1}{\alpha+2}=\frac{l_{\mbox{\tiny triple}}}{q}.
\end{equation}

As before, we assume that the source images are uniformly distributed at the
source plane inside radius $L_{\mbox{\tiny max}}$. Now the source image
coordinate limit (\ref{triplecond}) leads directly to a simple
graphical probability for triple image lensing
\begin{equation}\label{tripleprob}
P_{\mbox{\tiny triple}}(\alpha)=1-\frac{\alpha^2}{4}.
\end{equation}
that depends only on the cusp slope parameter of the density profile
(\ref{profile}). The expression (\ref{tripleprob}) for probability of
the triple image lensing can be considered also valid with reasonable
values of ellipticities for triaxial haloes. Corresponding treatment
can be made for haloes with elliptic isocontours following the profile
(\ref{profile}). Note, that the probability (\ref{tripleprob}) applies
only on proportional share of triple image lenses among
already identified lens systems, not on arbitrary galaxy
in the Universe. This is why the $q$ term dependency vanishes from the
equation. Additionally, this result does not account for the finite
angular resolution in the real world observations.

The remarkably simple condition (\ref{triplecond}) and probability
(\ref{tripleprob}) have direct consequences on the general
properties of the lensing potential and properties of dark matter
haloes. According to the
expression (\ref{tripleprob}),
double image lensing and triple image lensing have equal probability
when $\alpha=\alpha_{\mbox{\tiny eq}}=-\sqrt{2}\sim-1.4$. If the cusp slope value
$\alpha>\alpha_{\mbox{\tiny eq}}$, triple image lenses are more common than
double image lenses. The fact that there are only three
known cases with visible inner caustic image out of roughly hundred
implicates that the cusp slope in the dark matter profile must be closer to
isothermal value $\alpha\sim -2$ than $\alpha_{\mbox{\tiny eq}}$.

In fact, if we set $P_{\mbox{\tiny triple}}=N_{\mbox{\tiny triple}}/N=0.03$ and solve the cusp
slope parameter from the equation (\ref{tripleprob}), we get
$\alpha\approx-1.97$ - a result rather close to the quoted value for the
population H1 in this paper.

\subsection{Theoretical CSL probability}
\label{CSLdistro}

Here we show how to calculate a numerical probability
function for the CSL limit values. We assume that the source images
are uniformly distributed at the double image region of the source
plane limited by the equation (\ref{triplecond}), i.e. the radial
source coordinate
\begin{equation}
y=\frac{l}{qL_{\mbox{\tiny max}}}<-\frac{\alpha}{2}\le 1.
\end{equation}

We assume a single population of lenses with a universal cusp slope
value $\alpha$. All the slope parameters used in the following calculations
have $-2<\alpha<-1$. Special cases $\alpha=-1$ and $\alpha=-2$ could
be treated in similar way, but it is not necessary at this point.

The differential probability for having a lens with value
$\alpha_{\mbox{\tiny CSL}}$ is corresponding to an area of a ring with
an inner radius $y$ and an outer radius $y+\mbox{d}y$:
\begin{equation}\label{difpop}
\mbox{d}P=8\frac{y}{\alpha^2}\mbox{d}y,
\end{equation}
where $y=y(\alpha_{\mbox{\tiny CSL}};\alpha)$.
This equation can be integrated in order to estimate the probability of
a double image lens to have $\alpha_2<\alpha_{\mbox{\tiny CSL}}<\alpha_1$:
\begin{equation}\label{CSLprob}
P(\alpha_2<\alpha_{\mbox{\tiny CSL}}<\alpha_1)=4\frac{y^2(\alpha_2;\alpha)-y^2(\alpha_1;\alpha)}{\alpha^2}.
\end{equation}

Although it could be possible to write an approximation for the function
$y=y(\alpha_{\mbox{\tiny CSL}};\alpha)$ or rewrite the differential (\ref{difpop})
using other variables, we calculate it here numerically. For this
purpose we employ the magnification ratio $M=M(\alpha,\theta)$ of the
images $k_1$ and $k_2$ produced by the source image at $l$,
as presented in equation (\ref{opaxis}). Here
$\theta=k_1/k_2$ is the image coordinate ratio.

We start calculating the $y(\alpha_{\mbox{\tiny CSL}};\alpha)$ for each $\alpha_{\mbox{\tiny CSL}}$
by solving the equation
\begin{equation}\label{firsteq}
\frac{\mbox{d}M}{\mbox{d}\theta}_{{\big|}_{\tiny
\begin{array}{llll}
\alpha  =  \alpha_{\mbox{\tiny CSL}} \\
\theta  =  \theta_{\mbox{\tiny CSL}}
\end{array}}}  =  0
\end{equation}
numerically for $\theta_{\mbox{\tiny CSL}}$. The acquired value is subsequently inserted to
\begin{equation}\label{secondeq}
M(\alpha_{\mbox{\tiny CSL}},\theta_{\mbox{\tiny CSL}})=M(\alpha,\theta),
\end{equation}
which is then solved numerically for $\theta$. Equation (\ref{secondeq}) has
two solutions $\theta=\theta_{(u)}(\alpha_{\mbox{\tiny CSL}})$ and
$\theta=\theta_{(d)}(\alpha_{\mbox{\tiny CSL}})$.
When these solutions are inserted into the lens equation
\begin{equation}
y=y(\theta,\alpha)=
\frac{\theta}{L_{\mbox{\tiny max}}}\left[
\frac{1+\theta}{1+\theta^{\alpha+2}}\right]^{1/(\alpha+1)}
\frac{\theta^{\alpha+1}-1}{\theta^{\alpha+2}+1},
\end{equation}
source coordinate values $y_{(u)}(\alpha_{\mbox{\tiny CSL}};\alpha)$ and
$y_{(d)}(\alpha_{\mbox{\tiny CSL}};\alpha)$ are
acquired. Because we must limit the solutions to the double image cases,
resulting  $y_{(u)}$ or $y_{(d)}$ values are not allowed to exceed the limit
(\ref{triplecond}).

With these two solutions, the probability (\ref{CSLprob}) becomes
$$
P(\alpha_2<\alpha_{\mbox{\tiny CSL}}<\alpha_1)=
$$
\begin{equation}\label{probo}
4\frac{
y^2_{(d)}(\alpha_2;\alpha)-y^2_{(d)}(\alpha_1;\alpha)+
y^2_{(u)}(\alpha_2;\alpha)-y^2_{(u)}(\alpha_1;\alpha)}
{\alpha^2}.
\end{equation}

\section{Double Image Lens Systems}\label{data}

  
\begin{table*}
\vbox to220mm{\vfill\scriptsize
\caption{\label{obse} Double image lens systems used in the study.}
\rotatebox{90}{
\begin{tabular}{@{}lccllllllllllll@{}}
\hline
{System}                   &
{lens}                     &
{gal}                      &
{radio $\mu_1/\mu_2$}      &
{optical $\mu_1/\mu_2$}    &
{$z_{\mbox{\tiny L}}$}                    &
{$z_{\mbox{\tiny S}}$}        &
{radio $\alpha_{\mbox{\tiny CSL}}$} &
{optical $\alpha_{\mbox{\tiny CSL}}$} & 
{references} \\
\hline
HE0047-1756     &   &   &                     & $0.27 \pm 0.02$     & 0.41      & 1.66       &                     & $-1.848\pm 0.010$   & \cite{wiso2004}\\
HST01247+0352   & - &   &                     &                     &           &            &                     &                     & \cite{ratn1999}\\
HST01248+0351   & - &   &                     &                     &           &            &                     &                     & \cite{ratn1999}\\
Q0142-100       &   &   &                     & $0.122 \pm 0.003$   & 0.49      & 2.72       &                     & $-1.934\pm 0.002$   & \cite{surd1988,leha2000}\\
QJ0158-4325     &   &   &                     & $0.3477 \pm 0.0013$ &           & 1.29       &                     & $-1.7941\pm 0.0009$ & \cite{morg1999}\\
SDSS0246-0825   &   &   &                     & $0.322 \pm 0.007$   &           & 1.68       &                     & $-1.811\pm 0.004$   & \cite{inad2005}\\
CLASS B0445+123 & - &   & $0.152 \pm 0.005$   &                     & 0.557     &            & $-1.917\pm 0.003$   &                     & \cite{argo2003} \\
HE0512-3329     & - &   &                     & $0.90 \pm 0.07$     & (0.93)    & 1.57       &                     & $-1.24\pm 0.12$     & \cite{wuck2003}\\
CLASS B0631+519 &   & x & $0.129 \pm 0.005$   &                     & 0.09/0.62 &            & $-1.930\pm 0.003$   &                     & \cite{york2005} \\
CLASS B0739+366 &   &   & $0.171 \pm 0.011$   & $0.083 \pm 0.011$   &           &            & $-1.906\pm 0.006$   & $-1.956\pm 0.006$   & \cite{marl2001} \\
HS0818+1227     &   &   &                     & $0.145 \pm 0.010$   & 0.39      & 3.115      &                     & $-1.921\pm 0.006$   & \cite{hage2000}\\
CLASS B0827+525 & - &   & $0.374 \pm 0.002$   & $0.06 \pm 0.02$     &           & 3.87       & $-1.7758\pm 0.0015$ & $-1.967\pm 0.009$   & \cite{koop2000} \\
CLASS B0850+054 & - &   & $0.164 \pm 0.004$   &                     & 0.59      & 1.14/3.93? & $-1.910\pm 0.002$   &                     & \cite{bigg2003} \\
SDSS0903+5028   &   &   &                     & $0.477 \pm 0.012$   & 0.388     & 3.605      &                     & $-1.702\pm 0.009$   & \cite{john2003}\\
SBS0909+523     &   &   &                     & $0.80 \pm 0.03$     & 0.83      & 1.38       &                     & $-1.40\pm 0.03$     & \cite{leha2000}\\
RXJ0921+4529    &   &   &                     & $0.215 \pm 0.008$   & 0.31      & 1.65       &                     & $-1.880\pm 0.005$   & \cite{muno2001}\\
FBQ0951+2635    &   &   & $0.2143 \pm 0.0015$ & $0.306 \pm 0.010$   & (0.24)    & 1.24       & $-1.8799\pm 0.0010$ & $-1.822\pm 0.006$   & \cite{sche1998}\\
BRI0952-0115    &   &   &                     & $0.289 \pm 0.005$   & (0.41)    & 4.50       &                     & $-1.833\pm 0.003$   & \cite{leha2000}\\
Q0957+561       &   &   & $0.757 \pm 0.005$   & $0.94 \pm 0.03$     & 0.36      & 1.41       & $-1.449\pm 0.005$   & $-1.19\pm 0.06$     & \cite{gree1985}\\
SDSS1001+5027   &   & x &                     & $0.740 \pm 0.005$   &           & 1.84       &                     & $-1.466\pm 0.006$   & \cite{ogur2005}\\
J1004+1229      & - &   & $0.44 \pm 0.05$     & $0.23 \pm 0.04$     & 0.95      & 2.65       & $-1.73\pm 0.04$     & $-1.87\pm 0.02$     & \cite{lacy2002}\\
Q1017-207       &   &   &                     & $0.138 \pm 0.004$   & (0.78)    & 2.55       &                     & $-1.925\pm 0.002$   & \cite{surd1997,leha2000}\\
SDSS1021+4913   &   & x &                     & $0.40 \pm 0.03$     &           & 1.72       &                     & $-1.76\pm 0.02$     & \cite{pind2006}\\
JVAS B1030+074  &   &   & $0.0659 \pm 0.0014$ & $0.0280 \pm 0.0008$ & 0.60      & 1.54       & $-1.9654\pm 0.0008$ & $-1.9857\pm 0.0004$ & \cite{xant1998,leha2000}\\
HE1104-1805     &   &   &                     & $0.241 \pm 0.009$   & 0.73      & 2.32       &                     & $-1.863\pm 0.005$   & \cite{wiso1993,leha2000}\\
CLASS B1127+385 & - &   & $0.80 \pm 0.07$     &                     &           &            & $-1.39\pm 0.08$     &                     & \cite{koop1999} \\
CLASS B1152+199 &   & x & $0.3304 \pm 0.0012$ & $0.0225 \pm 0.0006$ & 0.439     & 1.019      & $-1.8058\pm 0.0008$ & $-1.9885\pm 0.0003$ & \cite{toft2000,rusi2002}\\
SDSS1155+6346   &   &   &                     & $0.48 \pm 0.06$     & 0.176     & 2.89       &                     & $-1.70\pm 0.05$     & \cite{pind2004}\\
SDSS1206+4332   &   & x &                     & $0.637 \pm 0.008$   &           & 1.79       &                     & $-1.568\pm 0.007$   & \cite{ogur2005}\\
Q1208+101       & - &   &                     & $0.229 \pm 0.007$   &           & 3.80       &                     & $-1.871\pm 0.004$   & \cite{maga1992,leha2000}\\
SDSS1226-0006   & - &   &                     &                     &           &            &                     &                     & \cite{ogur2004,eige2006}\\
SDSS1335+0118   &   &   &                     & $0.346 \pm 0.002$   &           &            &                     & $-1.7954\pm 0.0014$ & \cite{ogur2004}\\
Q1355-2257      &   &   &                     & $0.1934 \pm 0.0005$ &           & 2.00       &                     & $-1.8925\pm 0.0003$ & \cite{morg2003a}\\
HST14164+5215   & - &   &                     &                     &           &            &                     &                     & \cite{ratn1999}\\
CLASS B1600+434 &   &   & $0.80 \pm 0.03$     & $0.77 \pm 0.05$     & 0.41      & 1.59       & $-1.39\pm 0.03$     & $ -1.43\pm 0.05$    & \cite{jack2000}\\
PMNJ1632-0033   &   &   & $0.0765 \pm 0.0013$ & $0.031 \pm 0.003$   &           & 3.424      & $-1.9597\pm 0.0007$ & $-1.984\pm 0.002$   & \cite{winn2002}\\
FBQ1633+3134    &   &   &                     & $0.297 \pm 0.003$   &           & 1.52       &                     & $-1.828\pm 0.002$   & \cite{morg2001}\\
SDSS1650+4251   &   &   &                     & $0.144 \pm 0.010$   &           & 1.54       &                     & $-1.922\pm 0.006$   & \cite{morg2003b}\\
PKS1830-211     &   & x & $0.743 \pm 0.005$   & $0.0156 \pm 0.0015$ & 0.89      & 2.51       & $-1.464\pm 0.005$   & $-1.9921\pm 0.0008$ & \cite{subr1990,leha2000}\\
PMNJ1838-3427   &   &   & $0.0697 \pm 0.0015$ & $0.038 \pm 0.006$   &           & 2.78       & $-1.9634\pm 0.0008$ & $-1.981\pm 0.003$   & \cite{winn2000}\\
PMNJ2004-1349   &   & x & $0.98 \pm 0.03$     & $0.89 \pm 0.08$     &           &            & $-1.10\pm 0.07$     & $-1.25\pm 0.14$     & \cite{winn2001}\\
MG2016+112      &   & x & $1.06 \pm 0.05$     & $0.75 \pm 0.02$     & 1.01      & 3.27       &                     & $-1.45\pm 0.02$     & \cite{schn1985}\\
B2108+213       &   & x & $0.47 \pm 0.02$     & $0.39 \pm 0.05$     &           &            & $-1.71\pm 0.02$     & $-1.76\pm 0.03$     & \cite{mcke2005} \\
CLASS B2319+051 & - &   & $0.1979 \pm 0.0013$ &                     & 0.62      &            & $-1.8899\pm 0.0008$ &                     & \cite{rusi2001}\\
\hline
\end{tabular}
}
\vfil}
\end{table*}


For this study, we composed a catalogue of double image lens systems,
which have a configuration resembling axially symmetric lensing. The
catalogue is based on JVAS-CLASS radio survey and optical
CASTLES survey
and it
contains 44 lens systems, see table \ref{obse} and relevant
references. Four of these
systems had no data available during the conducted background
research, which reduces number of useful systems down to 40.

The Jodrell Bank - VLA Astrometric Survey
\citep[JVAS,][]{patn1992,brow1998,wilk1998}
  and the Cosmic Lens All Sky Survey
\citep[CLASS,][]{jack1995,myer1995}
  were originally geared towards finding
  new gravitationally lensed systems in order to determine the
  value of  the Hubble's constant $H_0$. During the surveys, they
  observed roughly 10000 flat spectrum radio sources. At the moment
  these surveys has detected 20 strong lens systems.
The Cfa-Arizona Space Telescope LEns Survey is an ongoing
  HST survey of all the known strong lenses and lens
  candidates. The main goal of the survey is to create an uniform high
quality photometric sample of these objects \citep{muno1998}.
\footnote{C.S. Kochanek, E.E. Falco, C. Impey, J. Lehar,
  B. McLeod, H.-W. Rix, the CASTLES survey has a website
  \texttt{http://www.cfa.harvard.edu/castles/}}.

Our subsample of these surveys was
chosen according to the following criteria: the lens
system must have two clearly separate lensed images corresponding to a
 quasar source. Triple and  quad lenses, Einstein rings and clearly
non-axially-symmetric cases were excluded.
Additional lensed components, such as radio jets were allowed as
long as the lensed point source was present.

Table \ref{obse} summarizes properties of the accepted lens
systems. After the name of the lens system, 
absence of a lens object candidate is indicated by a '-'
in the second
column of the table. If there are additional components at
the immediate surroundings of the lens system, it is indicated with
a 'x' in the third column.
All the available flux measurements from radio and optical
observations were used to calculate mean flux ratio weighed with
measurement errors in observations (columns four and five). The flux
ratio is calculated by dividing the closer (dimmer) image flux with the further
(brighter) image flux.
Columns six and seven contain evaluation for redshifts of the lens
candidate (if present) and the source correspondingly.
The eighth and ninth columns in the table
contain radio and optical cusp slope limit values
($\alpha_{\mbox{\tiny CSL}}$). Note, that these may have different
values because the radio and optical sources have different locations
at the source plane. The last
column lists the relevant references pointing to the source of the data
and/or the publication reporting the discovery of the lens system.

After the preliminary
Monte Carlo testing, we will use
the calculated CSL values in the final statistical analysis.
Because the optical data is more abundant, we have
chosen to prefer optical values, and use radio measurements only if the
optical data is not available.

\section{Monte Carlo testing}\label{probs}

  
\begin{figure*}
\center\includegraphics[width=1.0\linewidth]{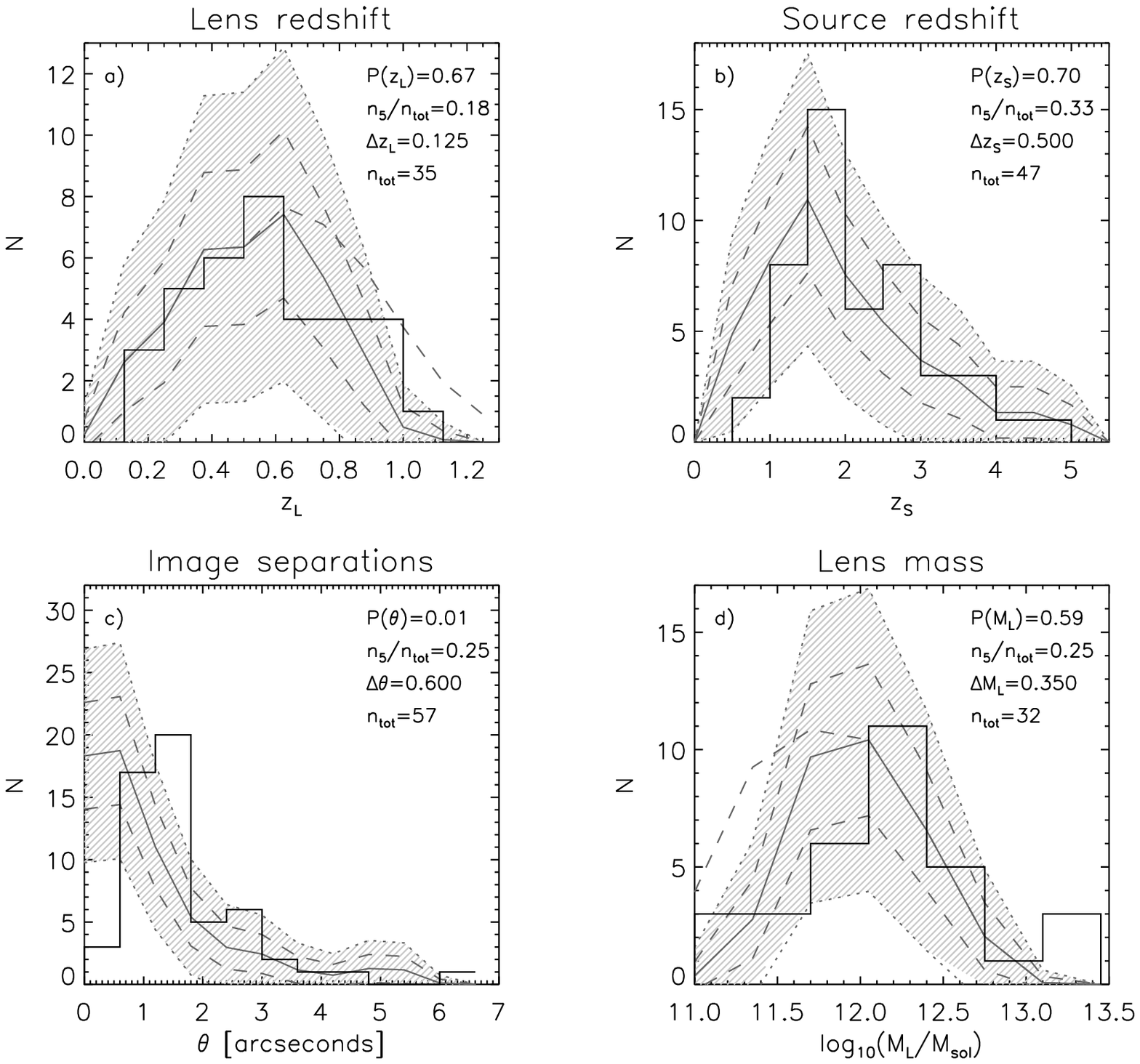}
\caption{\label{figmctest} 
Lens data generated by the simplest model in our Monte Carlo code. Panel a) presents
distribution of the lens redshifts. The sampled redshift distribution
contains some lenses with redshifts $z_{\mbox{\tiny L}}>1$ (dashed line), but they
vanish if the lens object magnitudes are accounted for (solid line).
Shaded region outlines $2\sigma$-region corresponding the Poisson
noise proportional to the observed lens sample size. Over-plotted
histogram describes redshift distribution of observed lenses.
Panel b) shows similar redshift distribution for the sources. The image
separations in arcseconds ($\theta$) are
presented in panel c) and sampled lens massed compared to the lens
masses inferred from observed magnitudes for the lens objects are
plotted in panel d). Note, that here the symbol $\theta$ 
for the image separation should not be confused
with the image coordinate ratio in equation
(\ref{opaxis}). The observed data is from the CASTLES survey catalogue. 
Each panel lists a Kolmogorov -- Smirnov test probability $P$ against
observed data, ratio of the bins with more than five data points ($n_5$)
to the total number of data points ($n_{\mbox{\tiny tot}}$), binsizes
($\Delta z_{\mbox{\tiny L}}$, $\Delta z_{\mbox{\tiny S}}$,
$\Delta\theta$, $\Delta M_{\mbox{\tiny L}}$), and total number of lens
systems ($n_{\mbox{\tiny tot}}$).
}
\end{figure*}
  

In order to test the CSL-analysis method against different
perturbations and investigate the error sources,
we have developed a computer code that generates mock lens
catalogues. Our code uses rejection method with base distributions
that are randomly sampled for initial redshifts, lens masses and
source luminosities.
Sampled values are fed to the lens equation solver that accounts for
the lens asymmetries and other more sophisticated properties of the
lensing event. Generated lens system is accepted or rejected according to
the observed configuration, selected magnitude limit and minimum
image separation threshold.

We use differential optical depth
$\frac{\mbox{d}\tau}{\mbox{d}z}(M_L,z_{\mbox{\tiny s}})$, as a
function of lens mass $M_{\mbox{\tiny L}}$ and source redshift
$z_{\mbox{\tiny S}}$ for the base lens distribution \citep{ofek2003}. Our
generator uses similar morphological mix for the lens galaxies, and
we have chosen parameter values $U=-0.20$, $P=1.20$ and
$f_{\mbox{\tiny E}}=0.95$ for the distribution. See \cite{ofek2003}
for further details, and information on other parameters affecting the
distribution.

The source objects are sampled from a double power law luminosity
function  $\Phi(L,z_{\mbox{\tiny S}})$ of quasars
as presented by \cite{wyit2002}. They use a luminosity function with pure
luminosity evolution and additional break for very high redshift
objects. See \cite{wyit2002} for further details.

Our parameters generate reasonably good samples of lens systems, as
can be seen from figure \ref{figmctest}. We compare our mock data
against observed redshift distribution of lenses,
redshift distribution of sources, image separations and
lens masses inferred from the
Faber--Jackson relation. We use Kolmogorov--Smirnov two-sample test
providing a probability that the
samples are drawn from the same background distribution.
When observed lens object magnitudes are converted to
absolute luminosities for mass estimation,
evolutionary and K-corrections are applied to the data
as presented by \cite{pogg1997} and \cite{bick2004}.
Note, that the size of the observational sample is changing in each
panel. The
reason for this is that we use all the double image lens
data available for testing, and some lens systems have missing data.

In figure \ref{figmctest}, the panel c) presenting image separation
distribution has low Kolmogorov --
Smirnov probability. Note, that here $\theta$ is image separation in
arcseconds, that should not be confused with the image coordinate
ratio in equation \ref{opaxis}. 
The mock data distribution follows the observed
image separations when $\theta\gtrsim 1$'', while lenses with tighter
image separations are over represented. We believe that
this discrepancy is a result from an observational bias. Because the
observed lens systems are not from a single systematic
survey, the lenses with small image separations are under-represented
in observations (lenses with image separation $\gtrsim 1$'' are easier to detect). 
It is clear, that the lens data acquired with a large array
of different instruments is far from being statistically
uniform. Therefore, chosen values for the magnitude limit and image
separation limit were determined from fitting to the data. 

Although we
conducted a parameter sweep for magnitude limit $M_{\mbox{\tiny lim}}$,
image separation
limit $\theta_{\mbox{\tiny lim}}$ and parameters $U$, $P$, $f_{\mbox{\tiny E}}$, it should be
emphasized that we do  not try to find the best fit for the
base-distribution parameters.  We use these values only for providing
as realistic
test data for the CSL-analysis method as possible.
It should also be noted that even though
the generated data is not
exactly fitting the observed sample, it should contain similar
internal correlations as the real data. Therefore it can provide
decent test cases for our method. We use bolometric
magnitude limit is $M_{\mbox{\tiny lim}}=30.5$ mag and image separation
limit $\theta_{\mbox{\tiny lim}}=0.2$''. Table 
\ref{para} summarizes chosen parameters for the base distributions.
For further explanation about physical
meaning of the parameters, see \cite{ofek2003} and
\cite{wyit2002}.

In the following subsections, we explore the effects from the
ellipticity of the lensing potential, the substructure within
the lensing halo, and the source variability coupled to the time-delay
that can distort the magnification ratios from the lensing
events. This is done by creating mock lens catalogues with different
degrees of perturbations and performing the CSL-analysis on them.


\begin{table}
\centering
\caption{\label{para} Numerical values for the parameters used by
  our Monte Carlo code. The upper part of the table lists parameters for
the Faber-Jackson relation and redshift evolution in the differential
optical depth for lensing $\mbox{d}\tau/\mbox{d}z(M_{\mbox{\tiny L}},z_{\mbox{\tiny S}})$.
The lower
part contains parameters for the quasar luminosity function $\Phi(L,z_{\mbox{\tiny s}})$. See text for the quoted references on the further details about the
parameters.}
\begin{tabular}{@{}llccc@{}}
\hline
{}            &
{parameter}   &
{Spiral}      &
{S0}          &
{Elliptical}  \\
\hline
$\frac{\mbox{d}\tau}{\mbox{d}z}$  & $\alpha$  &  -1.16 & -0.54  & -0.54\\ 
& $n_* (h^3\mbox{Mpc}^{-3})$      & $1.46\times 10^{-2}$ & $0.61\times 10^{-2}$ &
$0.39\times 10^{-2}$ \\
& $\gamma$   & 2.6 & 4.0 & 4.0 \\
& $\sigma_* (\mbox{km s}^{-1}) $ & 144 & 206 & 225 \\
& $U$ & -0.20 & -0.20 & -0.20 \\
& $P$ & 1.20 & 1.20 & 1.20 \\
& $f_{\mbox{\tiny E}}$ & 0.95 & 0.95 & 0.95\\ 

\\
& parameter & ($z_{\mbox{\tiny s}}> 3$) & ($z_{\mbox{\tiny s}}<3$) \\
\hline
$\Phi$ & $\beta_h$ & 2.58 & 3.43\\
& $\beta_l$ & 1.64 & 1.64 \\
& $\Phi_*(\mbox{Gpc}^{-1})$ & 624 & 624 \\
& $L_{*,0}(L_\odot)$ & $1.50\times 10^{11}$ & $1.50\times 10^{11}$\\
& $z_*$ & 1.60 & 1.60 \\
& $\zeta$ & 2.65 & 2.65 \\
& $\xi$ &3.30 & 3.30\\
\hline
\end{tabular}
\end{table}
                                                                              

\subsection{The Basic Concept}\label{basic}

The concept of the CSL value is to disentangle cosmology and
geometrical configuration of the each lens
event from the properties of the lens halo profile. These factors cancel
out in the derivation of the CSL equations (\ref{equation1}) and
(\ref{equation2}). However, they do have an indirect
influence on the underlying sample of the lens systems and can affect
the statistics of the CSL values.

The disentanglement of the lens halo properties from the
other factors in the CSL analysis is possible because
we assume simple lensing properties -- the lens potential is axially
symmetric and the source images are uniformly distributed on the strong
lensing region of the source plane. If there is a strong bias in the actual
source image locations on the source plane, it can distort the resulting CSL
distribution. Such a bias can emerge in a magnitude limited sample,
where sources at the high magnification region of the source plane are
over-represented in the total sample. 

\begin{figure}
\center\includegraphics[width=1.0\linewidth]{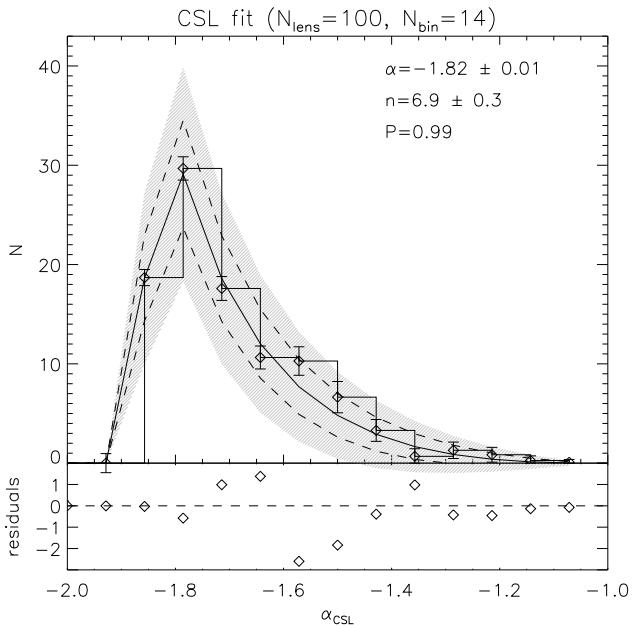}
\caption{\label{exafig}
Fit of a re-normalized theoretical probability function
for the CSL distribution of a lens population
with cusp slope $\alpha=-1.8$. Jagged histogram and corresponding
diamond symbols with error bars present histogram data for $\alpha_{\mbox{\tiny CSL}}$
values calculated from the mock sample. Error bars correspond to 0.1
magnitude absolute error in photometry. Thick solid curve represents fitted
CSL-distribution, with $\sigma$ (dashed line) and $2\sigma$ (shaded region)
outlining the Poisson shot noise error region. The fitted cusp slope value
for the data is $\alpha=-1.82\pm 0.01$ and
$n=6.9\pm 0.3$ with $P<0.01$ rejection probability.
}
\end{figure}

We have constructed a numerical probability function for
the CSL values ($\alpha_{\mbox{\tiny CSL}}$)
with an universal cusp slope $\alpha$ (see
appendix \ref{CSLdistro}). We fit this function to a histogram of
$\alpha_{\mbox{\tiny CSL}}$ values calculated from the (mock) data. In the fitting
procedure, we use the
Levenberg-Marquardt method and standard $\chi^2$-statistics.
The fitted parameters
are the universal cusp slope value $\alpha$ and normalization parameter $n$.
Figure \ref{exafig} illustrates such a fit.

\begin{figure}
\center\includegraphics[width=1.0\linewidth]{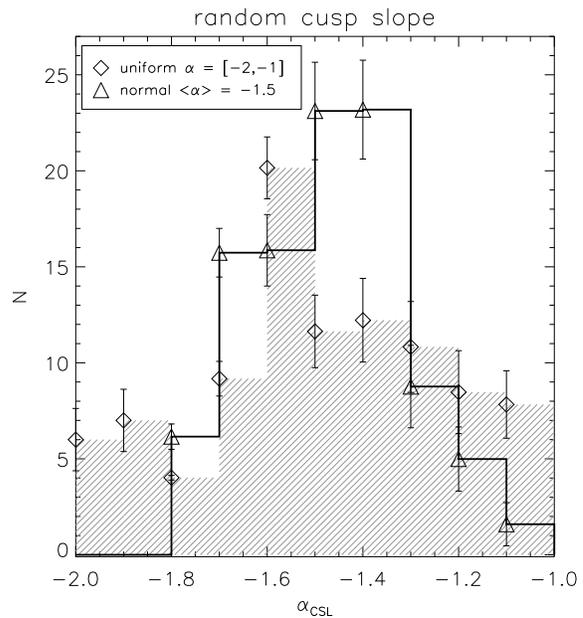}
\caption{\label{basic2}
If the cusp slopes of lensing haloes are random, characteristic
CSL-distribution (see figure \ref{exafig}) disappears. This
figure illustrates how a population of lenses with random cusp slope
values appears in the CSL histogram. The hatched area is a histogram of
a lens population with cusp slope values that have uniform
distribution. Thick line presents histogram of a lenses with normal
cusp slope values with an expectation value at
$\langle\alpha\rangle=-1.5$, and variance $\sigma_\alpha=0.15$.
Error bars represent 0.1 magnitude absolute error in photometry.
}
\end{figure}

The universal density profile has been one of the main assumptions in
studying the properties of dark matter haloes.
A population of lenses with a single cusp slope
parameter $\alpha$ produces a distinctive CSL-value distribution
(figure \ref{exafig}) that
is very different from  completely random profiles.
Figure \ref{basic2} shows CSL distribution of
a lens population with uniformly distributed random profiles with
$\alpha$ in range $[-2,-1]$ (hatched area), and second lens sample with
profile slope parameters in same range
with normal distribution
(expectation value $\langle\alpha\rangle=-1.5$ and
$\sigma_{\alpha}=0.15$, solid line).   

When performing fits on binned data, there are some issues to be considered,
see e.g. \cite{hump2009}. We tested several goodness-of-fit
functions and bin-sizes in order to find such a configuration that can
recover the universal cusp slope value from the data as often as
possible. We have chosen to use the previously mentioned
goodness-of-fit function and bin the data in 14 bins in range
$\alpha_{\mbox{\tiny CSL}}=[-2.0,-1.07]$. By using our binning selection and
goodness-of-fit function, we were able to recover the universal cusp
slope $\alpha$ with $2\sigma$ accuracy in most of the cases (cusp
slope value $\alpha=[-2.0,-1.1]$ and catalogue size $N_{\mbox{\tiny lens}}=[30,100]$).

The quoted error limits for the fitted parameters and corresponding
rejection probabilities are calculated with the Monte-Carlo method. We use
a bootstrap method to create hundred data sets that are fitted using the same
procedure as the original data. These fits are employed for error
estimates and rejection probability calculation. This is a well known
procedure that has been employed by several authors, see e.g. papers by
 \cite{buot2003} and \cite{hump2006} for further information.

There are certainly more suitable values for binning that can be
adapted for certain sizes of samples and values for the universal cusp slope
$\alpha$. However, we emphasize a good overall performance with
wide range of sizes for the lens samples and parameters.
If not mentioned otherwise, we use these choices in the analysis
presented in the following subsections.

  
\subsection{Lens Ellipticities}\label{ellisubs}

The CSL-analysis is based on an axially symmetric lens model.
In reality, the observations and the N-body
simulations indicate that the dark matter haloes are triaxial
ellipsoids, see e.g. \cite{haya2007} and references therein.
Here we examine the effects from asymmetries that can distort
the CSL-analysis results by changing the observed magnification ratios.

We have generalized our axially symmetric lens model to the elliptic
lens potential case. The isopotential surfaces
of the elliptic lens follow
similar surface density as the axially symmetric lens model, equation
(\ref{profile}). Our elliptic lens model approaches the axial
model with eccentricity value $e\rightarrow 0$. The lens
equations were derived from numerical integrals for elliptic
lensing presented in \cite{schr1990} and \cite{schn1992}.

The elliptic lenses can produce four image lenses, when the source
image is located inside the inner diamond shaped caustic curve. This
configuration is related to the Einstein ring in the case of axial
symmetry. Our Monte Carlo code accepts only double image lens systems
into the sample, i.e. all the source
images must be outside the inner caustic.
The size of the area at the source plane demarcated by the inner
caustic curve depends on the cusp slope $\alpha$ of the profile and
on the eccentricity $e$ of the halo. In general, a larger inner
caustic area means stronger deviations from the axially symmetric lensing.

\begin{figure}
\center\includegraphics[width=1.0\linewidth]{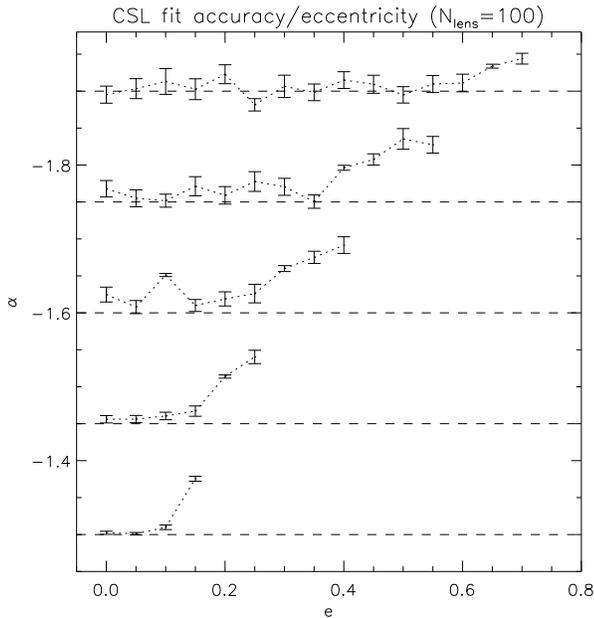}
\caption{\label{celfig}
We have compiled lens catalogues with cusp slope values
$\alpha=[-1.9,-1.75,-1.6,-1.45,-1.3]$ with constant
eccentricities. The lens catalogues have 100 lens systems. The x-axis
measures eccentricity $e$ and the y-axis cusp slope
$\alpha$. Each data point corresponds to a fitted CSL
distribution with error estimates.
All the fits with recognisable cutoff in the CSL value
distribution are plotted. If the eccentricity is further increased,
characteristic shape (figure \ref{exafig}) of the distribution vanishes.
}
\end{figure}

Strongly cusped lens profiles can tolerate high degrees of eccentricities
without
significant difference in the image flux ratio to the axially symmetric case.
Deviations from the axially symmetric model grow progressively larger
with shallower profiles and increasing eccentricities.
This trend is illustrated by figure
\ref{celfig}, in which we have examined an unrealistic scenario
where all the lenses have constant eccentricity. We have constructed
lens catalogues with cusp slope values
$\alpha=[-1.9,-1.75,-1.6,-1.45,-1.3]$ and varied eccentricity of the
lensing potential. We investigated eccentricities in range $e=[0,0.6]$
with 0.05 increments.

In figure \ref{celfig}, each cusp slope value is represented with a
horizontal dashed line. Plotted data points correspond to fits
to mock catalogues including error estimates with each value of
eccentricity. Data points are
plotted until the ellipticity makes the CSL distribution
unrecognisable (cutoff vanishes and tail becomes erratic). 
The figure clearly shows the aforementioned trend
for cusped profiles; the fit
tends to increasingly overestimate the cuspiness of the profile
with increasing halo ellipticity. The cutoff comes earlier with shallower
profiles.

We model the lens potential eccentricities with log-normal random
distribution with expectation value $\langle e \rangle$ and variance
$\sigma_e$. As a reference case, we follow the treatment by
\cite{hute2005} and adopt the distribution of ellipticities measured for
379 early-type galaxies by \cite{jorg1995}.
The distribution has mean $\langle e\rangle =0.31$ and $\sigma_e=0.18$,
with upper limit $e<0.6$.

\begin{figure}
\center\includegraphics[width=1.0\linewidth]{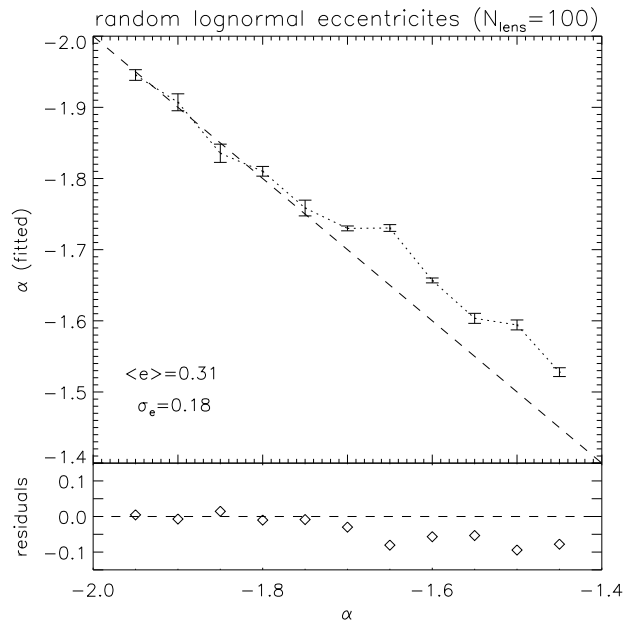}
\caption{\label{relfig}
Cusp slopes from the fitted CSL values using log-normal random distribution
for the lens halo eccentricities (see text). The x-axis has the cusp
slope value used in the mock catalogue generator, and the y-axis has
fitted cusp slope value. Diagonal dashed line marks the exact fit.
Each data point denotes a result from a CSL distribution fit to a
compiled lens catalogue, with error estimates.
Lower panes shows fit residuals. Fit gets
progressively worse with more shallow profile
until at $\alpha\sim -1.45$ the characteristic
shape of the CSL value distribution is too distorted in order to make
a sensible fit to the data.
}
\end{figure}

Our results indicate that, regardless of realistically elliptical lens
haloes, the modelled CSL value distribution can recover the universal cusp slope
up to slope value $\alpha\sim -1.7$ and give a rough guess for the
exact value up to $\alpha\sim -1.5$ that is always an overestimate.
See figure \ref{relfig}.

  
\subsection{Substructure}\label{subssubs}

The dark matter
haloes should have several sub-haloes that generate irregularities in
the lensing potential. These irregularities can corrupt the macro-lens
magnification ratio and distort the
CSL-distribution. Here we have chosen to follow
similar approach as in \cite{rozo2006}. We
model the effects from the substructure on the magnification of images
with $N$ linearized perturbers contributing with
$\delta\mu_i$ to the total perturbation
$\delta$. When contribution from all the perturbers is added together, we get
the total perturbation 
\begin{equation}\label{prtbsum}
\delta=\frac{\sum_i^N\delta\mu_i}{|\mu|}
=2|\mu|\left[(1-\kappa)\sum\limits_i^N\delta\kappa_i-
\gamma\sum\limits_i^N\delta\gamma_i\,\cos(2\phi_i)
\right],
\end{equation}
where $\kappa$ and $\gamma$ are  macro-lens convergence and shear,
whereas $|\mu|$ is macro-lens magnification.
Each perturber is located
at radial coordinates $\theta_i$ and $\phi_i$, and 
convergence perturbations $\delta\kappa_i$ and shear perturbations
$\delta\gamma_i$ are corresponding
contributions from the $i$th perturber. 
It is assumed that $|\delta\mu_i|\ll 1$,
thus the astrometric perturbations by the substructure
are negligible and 
only the changes in the image fluxes
are considered.

We follow the analysis by  \cite{dala2002} and \cite{rozo2006}
by employing a pseudo-Jaffe density profile
\begin{equation}
\rho(r)\propto r^{-2}(r^2+a^2)^{-1}
\end{equation}
for individual perturbers. Here
$r$ is radial coordinate and $a$ is an effective tidal radius.
As presented in \cite{rozo2006}, we parametrize
the tidal radius with $\lambda_a$ as
\begin{equation}
a=\frac{\pi\lambda_a\sqrt{b b_{\mbox{\tiny H}}}}{2},
\end{equation}
where $b$ is the Einstein radius of the perturber and $b_{\mbox{\tiny H}}$ the
corresponding radius for the macro-lens model.
Convergence and shear contribution ($\delta\kappa_i$ and $\delta\gamma_i$)
from the $i$th
perturber at polar coordinates $(\theta_i=k_i x_{\mbox{\tiny e}}
r_{\mbox{\tiny s}}/D_{\mbox{\tiny L}},\phi_i)$ is
\begin{equation}
\delta\kappa_i=\tilde b\left(\frac{1}{2\theta_i}-\frac{1}{2\xi}\right)
\end{equation}
and
\begin{equation}
\delta\gamma_i=\tilde b\left[\frac{1}{2\theta_i}+\frac{1}{2\xi}
-\frac{a}{\theta^2_i}\left(\frac\xi a-1\right)\right].
\end{equation} 
The mass of the perturber is $m=\pi a\tilde b\Sigma_{\mbox{\tiny cr}}D_{\mbox{\tiny L}}^2$, while
 $\tilde b$ is defined by
\begin{equation}
\frac{b}{\tilde b}=1+\frac ab-\left[1+\left(\frac ab\right)^2\right]
\end{equation}
and $\xi$ is
\begin{equation}
\xi\equiv\sqrt{\theta_i^2+a^2}. 
\end{equation}
When $a\rightarrow\infty$ the pseudo-Jaffe
  lens profile approaches SIS profile, and when $a\rightarrow 0$ the
  profile corresponds to a point mass lens.

Mass spectrum $s$ of the perturbers is assumed to follow
$\mbox{d}s/\mbox{d}m\propto m^\beta$, where we choose
$\beta=-1.8$ as in \cite{gao2004}. 
The amount of substructure is defined by a ratio of the substructure surface
density to the critical density $f_{\mbox{\tiny
    sub}}=2\Sigma_{\mbox{\tiny s}}/\Sigma_{\mbox{\tiny cr}}$. The mass
spectrum is parametrized also by the
cutoff at the substructure mass value $m_{\mbox{\tiny max}}$ that is
chosen to be 1 per cent of the mass within the Einstein radius
(mass $M_{\mbox{\tiny E}}$) of the lens
object.

  
\begin{figure}
\center\includegraphics[width=1.0\linewidth]{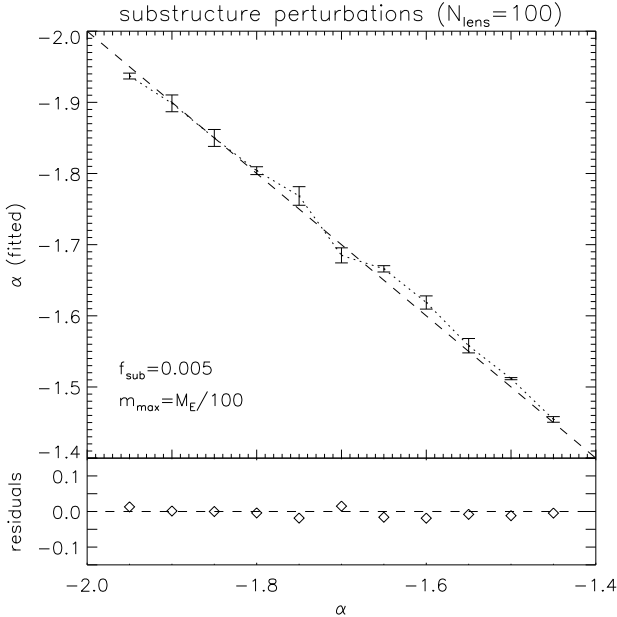}
\caption{\label{sbstfig}
Cusp slopes from the fitted CSL values including substructure
perturbations to then magnification. The x-axis has the cusp
slope value used in the mock catalogue generator, and y-axis has
fitted cups slope value. Diagonal dashed line marks the exact fit.
Each data point denotes a result from a CSL distribution fit to a
compiled lens catalogue, with error estimates.
Lower panes shows fit residuals.
}
\end{figure}
  

We account the substructure perturbations to the image magnification
in the synthetic lens catalogue generation as follows:
the magnification change of the each image is estimated
by randomly generating 1000 perturbers uniformly distributed at the
lens surface with a surface density that is determined by the
$f_{\mbox{\tiny sub}}$. The total perturbation $\delta_{tot}$ is then estimated
from the equation (\ref{prtbsum}) with negative
surface mass component in order to guarantee the mass conservation
\begin{equation}
\delta_{tot}=\delta-|\mu|(1-\kappa)f_{\mbox{\tiny sub}}.
\end{equation}
The perturber masses follow the previously discussed
power law mass spectrum with index $\beta$.
The total perturbation $\delta_{tot}$  is
then added to the total magnification of the image, corresponding
to the macro lens shear and convergence. 

Again, we follow the \cite{rozo2006} with the parametrization of the
substructure and choose tidal radius parameter $\lambda_a=4.0$, and
substructure surface density $f_{\mbox{\tiny sub}}=0.005$.

We study the goodness of the CSL fit as a function of the universal
cusp slope of the generated lens catalogue with constant strength for
the substructure. The lens catalogues are generated with an axially
symmetric lens model with cusp slopes $\alpha$ ranging in
$[-1.95,-1.45]$. The results are summarized in figure \ref{sbstfig}.
All the fits are very close to the original value, although it would
seem that fits get slightly worse at higher values for the cusp slope
($\alpha\gtrsim -1.7$).

Although the substructure can significantly alter the observed flux
ratio of the images, this can happen only at the vicinity of the
critical curves for lensing
(at the source coordinate
$l\sim 0$ or $l\sim l_{\mbox{\tiny max}}$).
These kinds of lens systems produce either strongly magnified Einstein
rings or triple image lenses that are excluded from
the analysis by the selection criteria. Furthermore,
because such lens systems
have incidence probability proportional to the square of the ratio
$l/l_{\mbox{\tiny max}}$, they should be comparatively rare
when the source images are
uniformly distributed at the source plane.

Here it should be noted, that a similar treatment could be applied for
micro-lensing.  However, because the
CSL-limit analysis is based on double image lenses where the images
are generally not close to the critical curves, we assume that the effects from
 micro-lensing can be safely neglected.

\subsection{Time-delay}\label{svrbsubs}

In the double image lenses, the lensed images represent the source at
different moments in time in the rest frame of the source. This is a
consequence of a different light propagation distance for each image
(geometric time-delay) and different gravitational potential along the
light ray trajectory (potential time delay).
As a consequence, the observed flux ratio is
different from the magnification ratio of the images if the source
flux is varying at suitable timescale. This can distort
the acquired CSL-distribution. 

We calculate the time delay from the lensing potential generated by
our parametrized density profile. For definitions of time delay, see
for example \cite{schn1992}. The source variability is modelled by
generation power law noise, that is parameterized according to the
observational relations published in \cite{vanb2004}.

Source flux variations are modelled by generating power law variations
$\Delta m$ around the baseline magnitude of the source sampled from
the quasar luminosity function. The power spectrum of
the magnitude variations is
\begin{equation}\label{pwrspec}
\Phi=\Phi_0\omega^{-\eta},
\end{equation}
where $\omega$ is frequency, $\eta$ power law index and $\Phi_0$
corresponding normalization.On generating power law noise,
see paper by \cite{timm1995}.

Noise with characteristics defined by (\ref{pwrspec}) have variance
\begin{equation}\label{noise}
\sigma^2_{\Delta m}=2\int\limits^\infty_{2\pi
  f}\mbox{d}\omega\Phi(\omega)=\frac{2\Phi_0(2\pi)^{-\eta}}{\eta-1}f^{1-\eta},
\end{equation}
where $f$ is the minimum frequency to be considered.
Corresponding covariance function for the magnitude $m=m(\tau)$ is
$$
B(\Delta\tau)=\langle m(\tau)m(\tau+\Delta\tau)\rangle=\langle\Delta m^2\rangle
$$
\begin{equation}\label{cova}
=\int\limits^\infty_{2\pi f}\mbox{d}\omega\cos(\omega\tau)\Phi(\omega)
=\frac 12\Phi_0\tau^{\eta-1}A(\eta,f),
\end{equation}
that is written at time $\tau$ with time lag $\Delta\tau$.
In previous equation we define $A(\eta,f)$ as
$$
A(\eta,f)=
$$
\begin{equation}
e^{-i\pi(1-\eta)/2}\Gamma(1-\eta,i2\pi f)+
e^{i\pi(1-\eta)/2}\Gamma(1-\eta,-i2\pi f),
\end{equation}
where $\Gamma(x,y)$ is the incomplete gamma-function. The variability
function $V=V(\tau)$ and model for it are defined as
\begin{equation}\label{strucfun}
V(\tau)=\sqrt{\frac\pi 2\langle\Delta m\rangle^2-\langle\sigma^2_{\mbox{\tiny s/n}}\rangle}
=V_0\left(\frac{\tau}{\tau_0}\right)^\gamma,
\end{equation}
where $V_0$ and $\tau_0$, that are determined from the observations,
parametrize the model. Here $\langle\sigma^2_{\mbox{\tiny s/n}}\rangle$
is the signal to noise ratio of the variations.

When $\langle\Delta m^2\rangle\sim \pi\langle\Delta m\rangle^2/2$, we get from the definition
of the variance and equations (\ref{noise}) and (\ref{cova}) the
relation between power spectrum parameters and the variability function.
The power law indices are related as
\begin{equation}
\eta=1+2\gamma
\end{equation}
and normalization as
\begin{equation}
\Phi_0=\frac{4V_0}{\pi A(\eta,f) \tau_0^{2\gamma}}.
\end{equation}
Here it should also
hold $\langle\sigma^2_{\mbox{\tiny s/n}}\rangle=\pi|\sigma^2_{\Delta m}|$.

The normalization $\Phi_0$  and the power law index $\eta$ 
characterizing the power law noise are calculated from
the variability function presented by \cite{vanb2004}, where they study
photometric variability of roughly 25 000 quasars in the SDSS survey.

In their paper, they estimate the variability function
power law index as $\gamma=0.246\pm 0.008$ and characteristic time
scale $\tau_0=(5.36\pm 1.46)\times 10^5$d. The normalization of the
variability function (\ref{strucfun}) is parametrized as
\begin{equation}
V_0=v(M)v(\lambda_{\mbox{\tiny R}})v(z)
\end{equation}
with the absolute magnitude of the quasar $M$,
observed rest-frame wavelength $\lambda_{\mbox{\tiny R}}$ and redshift of the quasar $z$.
Here we set
\begin{equation}
v(M)=10^{\beta M/2.5}
\end{equation}
in which $\beta=0.246\pm 0.005$. Correspondingly
\begin{equation}
v(\lambda_{\mbox{\tiny R}})=a_0e^{-\lambda/\lambda_0}+a_1
\end{equation}
with $a_0=0.616\pm0.056$, $a_1=0.164\pm 0.003$ and $\lambda_0=988\pm
60$\AA. And finally
\begin{equation}
v(z)=(0.019\pm0.002)z+(0.037\pm0.005).
\end{equation}

  
\begin{figure}
\center\includegraphics[width=1.0\linewidth]{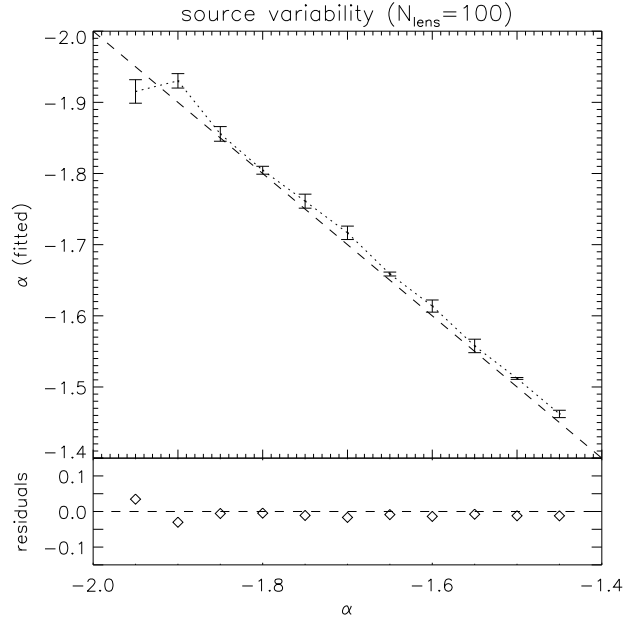}
\caption{\label{tmdlfig}
Cusp slopes from the fitted CSL values including source variability
induced time delay perturbations to the image flux. The x-axis has the cusp
slope value used in the mock catalogue generator, and y-axis has
fitted cusp slope value. Diagonal dashed line marks the exact fit.
Each data point denotes a result from a CSL distribution fit to a
compiled lens catalogue, including error estimates.
Lower panel shows fit residuals.
}
\end{figure}
  

The time lag $\Delta\tau$ is acquired from the time delay equations,
and we choose rest-frame wavelength corresponding to the observed
\emph{B}-band wavelength. The absolute magnitude is acquired from the sampled
luminosity of the quasar.
 
We convert the parametrized variability function to the power law
noise index $\eta$ and corresponding normalization $\Phi_0$,
and obtain random
change in the flux of the source by generating similar power law noise. 
This changed flux (that equals the baseline flux sampled from the
quasar luminosity function with the random change
for the each image) is then correspondingly magnified
by the lens and inserted into
the standard mock lens catalogue generator.

We study the effects from the time delay coupled to the source
variability by generating several mock catalogues with an universal
cusp slope values $\alpha$ ranging in  $[-1.95,-1.45]$. The mock catalogues are
generated with an axially symmetric model. The fitted CSL values are
presented in figure \ref{tmdlfig}.

Inclusion of the source variability seems to slightly strengthen
the acquired cusp from the lensing data. Although all the fits
are reasonably good, strongly cusped catalogues are most affected
by the source variability.

\subsection{Combined error sources}

In the previous subsections, we have examined the strength of several
distortions on the CSL statistics with the
simplest possible model.
Here we add all the contributing factors (ellipticities, halo
substructure and time-delay) together in order
to see what is the applicability of our method.

We study the goodness of the CSL fit as a function of the universal
cusp slope value in the mock catalogues. We use elliptical lens model with
random log-normal eccentricities as in subsection \ref{ellisubs}. The
modelled lens systems possess similarly parametrized substructure
as in \ref{subssubs} and source variability corresponding to the
previous subsection \ref{svrbsubs}.

\begin{figure}
\center\includegraphics[width=1.0\linewidth]{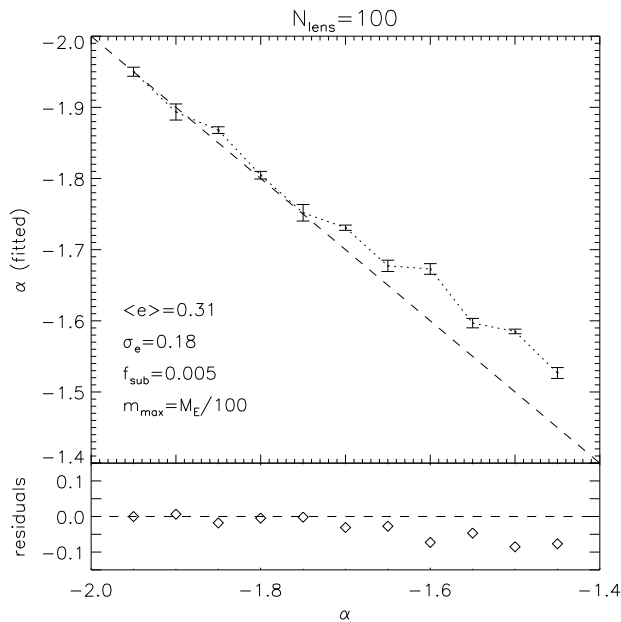}
\caption{\label{smryfig}
Cusp slopes from the fitted CSL values with all the examined
perturbation sources (lens ellipticity, substructure and time delay
effects coupled to the source variability). The x-axis has the cusp
slope value used in the mock catalogue generator, and y-axis has
fitted cusp slope value. Diagonal dashed line marks the exact fit.
Each data point denotes a result from a CSL distribution fit to a
compiled lens catalogue, including  error estimates.
Lower panel shows fit residuals.
}
\end{figure}

The results of the fits are presented in figure \ref{smryfig}. It would
seem that a simple axially symmetric model with an assumption of uniform
source image distribution at the source plane can capture essential
properties of the double image lenses in the presence of higher order
effects when calculating the CSL statistics.

In general,the CSL goodness-of-fit $\chi^2$-surface contains
a local minimum inside each bin. In the fits that are presented here,
we have always chosen the global minimum. However, in some cases there
are two local minima that have almost the same $\chi^2$ values resulting two
equally good fits in statistical sense. Such problems can be
resolved by manually adjusting the binning to cover the range of
$\alpha_{\mbox{\tiny CSL}}$ values in an optimal way. We have chosen to use a
fixed binning that produces fits that should
become progressively poorer
when the universal cusp slope approaches $\alpha=-1$ because
  the data is assigned to fewer bins correspondingly. However, the
  leftover zero bins are fitted into the model with estimated
  error bars, thus the recovered error bars for $\alpha$
  do not change significantly. This is a
conscious choice that was made in order to study the error
  sources as a function of the cusp slope
in a consistent way. In principle, an adaptive binning should
  produce better results, but as the real data is spread over rather
  wide range of the $\alpha_{\tiny\mbox{CSL}}$ values, we hold to
  our previously adopted binning scheme.

  
\section{Fitted CSL Statistics}

In this section, we 
apply the statistical analysis of the CSL values to
the real double image lens data.
We use our double image lens sample described in section \ref{data}
to study the distribution of the CSL values.
The calculated CSL values for each lens system
are presented in the table \ref{obse}.
These values are fitted to the theoretical distribution
(see appendix \ref{CSLdistro}). The fit produces estimates
for an universal cusp slope value $\alpha$ and corresponding
normalization $n$ proportional to the size of the sample.

In the fitting procedure, the error limits for the histogram 
were created using the Monte-Carlo method. Thousand distributions
were constructed with simulated Gaussian errors according to the
measurement errors in the flux ratios. The data points in the
presented histogram are mean values, and
the corresponding error limits are the standard deviation calculated
from these constructed histograms.
In some cases values at the end points of the distribution had zero error
limits. Those were estimated upwards with mean of non zero
errors within the distribution. We use the same setup for the data as for
the Monte Carlo testing described in the previous section.
The details are presented in section \ref{basic}.

\subsection{Single population model}

\begin{figure}
\center\includegraphics[width=1.0\linewidth]{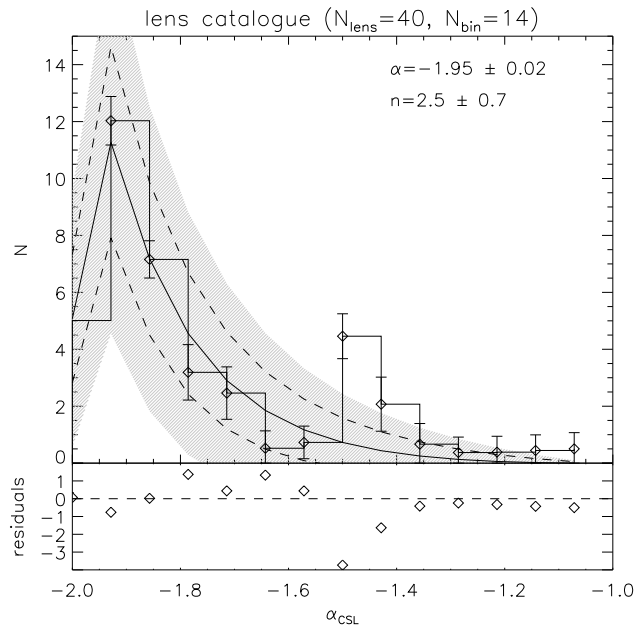}
\caption{\label{datfig1}
A single population fit 
of a re-normalized theoretical probability function
for the CSL value distribution from an observational lens catalogue.
Jagged histogram and corresponding
diamond symbols with error bars present histogram data for $\alpha_{\mbox{\tiny CSL}}$
values calculated from the data. Error bars correspond to measurement 
errors in photometry. Thick solid curve represents fitted
CSL-distribution, with $\sigma$ (dashed line) and $2\sigma$ (shaded region)
outlining the Poisson shot noise error region. The fitted cusp slope value
for the data is $\alpha=-1.95\pm 0.02$ and normalization
$n=2.5\pm 0.7$.
}
\end{figure}

We start our analysis with an assumption of a single universal halo
profile. The acquired cusp slope value for this lens population
is $\alpha=-1.95\pm 0.02$. The value of the normalization parameter is
not interesting, because it is simply scaling the probability function
to the size of the sample.
The resulting fit is presented in figure \ref{datfig1}. The acquired value
for the cusp is close to the value of an isothermal sphere (SIS).

All the data points below $\alpha_{\mbox{\tiny CSL}}\sim-1.5$ fall well into the
one Poisson-$\sigma$ range of the model. However, at
$\alpha_{\mbox{\tiny CSL}}>-1.5$ there are two data points well outside of the
$2\sigma$-Poisson noise range. These data points represent
roughly 18 per cent of the total sample of lens systems.
All corresponding lens systems cannot belong to the same profile group as the
acquired fit because perturbations have a tendency to steepen
the acquired cusp slope. One explanation could be
that these are strongly sheared systems, but there are several
arguments against this possibility.

The fact that the deviating lens systems are rather tightly
clustered around value $\alpha\sim -1.5$ does not support sheared
lens scenario. Randomly sheared systems should produce wider distribution,
with similar characteristics as random cusp slope values (see figure
\ref{basic2}). Strongly cusped haloes are also more resistant against
the deviations from the axial symmetry than the haloes with shallower cusps
(see section \ref{ellisubs}).
Additionally, sheared systems that have strongly deviating
magnification ratio from the axially symmetric value, very often produce
more than two images in similar fashion as strongly elliptic haloes.
Such lens systems are excluded from the analysis by the
selection criteria used to construct our lens sample.

A single universal cusp slope value cannot fully explain
the observed CSL-value distribution. Therefore, we investigate also a
possibility that this deviation is a signature of a second population
of haloes with their own characteristic density profile.

  
\begin{figure}
\center\includegraphics[width=1.0\linewidth]{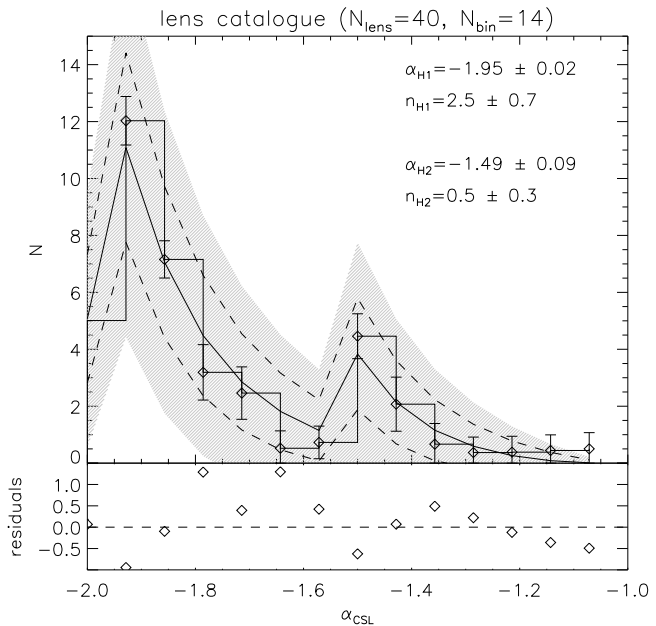}
\caption{\label{datfig2}
A dual population fit 
of a re-normalized theoretical probability function
for the CSL value distribution from an observational lens catalogue.
Jagged histogram and corresponding
diamond symbols with error bars present histogram data for $\alpha_{\mbox{\tiny CSL}}$
values calculated from the data. Error bars correspond to measurement 
errors in photometry. Thick solid curve represents fitted
CSL-distribution, with $\sigma$ (dashed line) and $2\sigma$ (shaded region)
outlining the Poisson shot noise error region. The fitted cusp slope value
for the data is $\alpha_{\tiny\mbox{H1}}=-1.95\pm 0.02$ and normalization
$n_{\tiny\mbox{H1}}=2.5\pm 0.7$
for the first population. The second population has
corresponding determined values $\alpha_{\mbox{\tiny H2}}=-1.49\pm 0.09$ and
$n_{\mbox{\tiny H2}}=0.5\pm 0.3$.
}
\end{figure}
  

\subsection{Dual population model}

The dual population model (populations are abbreviated as H1 and H2)
is implemented by adding two individually
normalized distributions together, producing a model with four
parameters. This model
is fitted to the observational data in a similar fashion as the single
population model in the previous subsection. The fit produces estimates
for the universal cusp slope values 
$\alpha_{\mbox{\tiny H1}}$ and $\alpha_{\mbox{\tiny H2}}$ for both populations,
and the corresponding normalization values $n_{\tiny\mbox{H1}}$ and
$n_{\tiny\mbox{H2}}$.

The detailed fit to this data
is presented in figure \ref{datfig2}. The acquired values are
cusp slope $\alpha_{\tiny\mbox{H1}}=-1.95\pm 0.02$
and normalization $n_{\tiny\mbox{H1}}=2.5\pm 0.7$
for the population H1 and for the population H2 correspondingly
$\alpha_{\mbox{\tiny H2}}=-1.49\pm 0.09$, $n_{\mbox{\tiny H2}}=0.5\pm 0.3$.
Normalization values $n_{\tiny\mbox{H1}}$ and  $n_{\tiny\mbox{H2}}$
gives the relative abundance
of population H1 and population H2 lens objects in the
sample. Our results indicate that 83 per cent of the lens objects belong to
the population H1 and 17 per cent to the population H2 with 4 per cent error
margins in both values. This suggests that on average roughly one sixth of the
lens objects capable of strong lensing belong to the second
population with shallower cusp slope.

The dual population fit is better than the single population fit,
which is to be expected because number of free parameters is
doubled. In practice all the data points are now inside the one-$\sigma$
Poisson noise range of the model.
Population I parameters are the same as in the single population
model. However, the second population parameters have much
larger relative error range than population H1 parameters,
indicating large uncertainty. This reflects the small size of the lens
sample. Additionally, the second population cusp
slope has such a value that is much more affected by the perturbations 
than the cusp slope value for population H1 lenses.

\subsection{Is the second population real?}

Regardless of the Poisson noise limits, we are very concerned if
the observed feature at $\alpha_{\mbox{\tiny CSL}}\sim -1.5$ is nonetheless noise induced.
The statistical fit with such a small sample compared to the number
of fitted parameters should be regarded with high suspicion. Additionally,
the signature of the second population is relatively weak when
considering the error limits for the fitted parameters of the second
population.

Therefore, we set up a Monte-Carlo experiment in order to find out the
probability that the second population hypothesis could be rejected.
We generated 100 lens catalogues with 40 lens systems by randomly
sampling a mock catalogue of 4000 lens systems.
The mock catalogue was created using a single lens population
with a cusp slope value of $\alpha=-1.95$ and accounting for
the lens ellipticities, the substructure and the source variability.
These generated catalogues were fitted with a single population model.
The probability to reject the second population hypothesis
is equal to the probability to get as strong signature of the
second population as in the observed lens catalogue.

In the observations, the second population feature exceeds the
$2\sigma$ noise limit by factor 1.8 at
$\alpha_{\mbox{\tiny CSL}}$ range $[-1.6,-1.4]$.
In order to interpret the noise induced feature as the second
population, we set a condition that the data-set
must contain a data point that exceeds the $2\sigma$ level by similar
factor at the same range. We found one such catalogue out of one hundred,
thus we estimate that the probability for rejection of the second
population hypothesis is at order of one per cent.

On the other hand, the observational lens data contains two data points
above the $2\sigma$ noise level. If we count lens samples that have
two data points exceeding the $2\sigma$ threshold in 
$\alpha_{\mbox{\tiny CSL}}$ range $[-1.6,-1.4]$, we get a rejection probability of two per cent. If
we add the factor of 1.8 excess condition, the probability goes down
below one per cent.

Hence, we estimate that the rejection probability for the dual
population hypothesis is roughly $\sim 1$ per cent.

\section{Discussion and conclusions}

Our analysis is based on a geometric measure of a lens
system - how much cuspiness does an axially symmetric lens need in
order to produce observed magnification ratio, i.e. the CSL value.
In a sense, analogously to the curvature characterizing properties of space,
it can be thought as a gauge characterizing localized
properties of a lensing event.
There is no need to define the exact global properties of
the lensing potential in order to have a meaningful measure.

The CSL measure can be derived for axially symmetric lenses, and
as a result, all the coefficients containing information on cosmology,
cosmological angular separations of the lens and the source, lens mass
and concentration are cancelled out. Only preserved quantities
are the cusp slope and the image magnification ratio.

If we assume that the observed lenses have completely random
orientations, thus neglecting effects from the magnification bias, a
theoretical distribution of the CSL values can be calculated for a
population of lenses. In that case, the properties of the lensing
profile(s) determine the shape of the distribution.
When we calculate the CSL value distributions from the real life
data, we make an
additional assumption that the observed image fluxes are determined
only by the lens magnification of a constant flux source.

This chain of assumptions has its weak points, and we have tried to
address most of them in section \ref{probs}. Here
we are going to do a brief summary of these, and discuss few
unaddressed ones.

The first unaddressed issue is at the derivation of the magnification
ratio equation (\ref{opaxis}). The equation is based on the inner part
of the lens equation that does not behave correctly when the radial
image coordinate at the lens plane exceeds the value $k_{\mbox{\tiny B}}$.
Usually, this happens
outside the strong lensing region, but if $\alpha\lesssim -2.0$ and
the mass of the lens or the concentration is high enough, $k_{\mbox{\tiny B}}$ can be located
inside the strong lensing area ($k_{\mbox{\tiny B}}<k_{\mbox{\tiny cr2}}$).

This asymptotic behaviour of the lens equation was accounted
for when the mock catalogues were created with the Monte Carlo method.
The method works correctly with the mock data because
the deviation from the correct
value at $k\gtrsim k_{\mbox{\tiny B}}$ happens very slowly with the increasing
radial image coordinate. Usually, difference becomes
non-negligible only for images far into the
weak lensing region even if $k_{\mbox{\tiny B}}$
were inside the strong lensing area. Significant deviation requires
unrealistically high mass or concentration for the halo.

The next issue is the error factors considered in section
\ref{probs} -- lens ellipticity, lens potential substructure, and source
variability. Our earlier treatment was focused 
on examining the perturbations in the recovered cusp
slope value. The analysis showed that the most severe disturbations
originated from the lens ellipticities, although those were at manageable
level when $\alpha<-1.4$ with random ellipticities.

The aforementioned factors have little or no effect because the
analysis is restricted to the double image lenses by definition.
At the double image region, deviations from the axially symmetric
model are at their weakest.
Differences become severe nearby the critical curves (at
the origin $l\sim 0$ and at the border of the strong lensing region
$l\sim l_{\mbox{\tiny max}}$), where the magnification diverges.
Images nearby the critical curves produce either the Einstein
rings ($l\sim 0$), quad lenses,
or triple image lenses ($l\lesssim l_{\mbox{\tiny max}}$).
Hence, the real life lensed images nearby critical curves are easily
identified, and they can be excluded from the observational sample.

The similar reasoning can be used to counter arguments for
effects from the sheared potential, micro-lensing,
and the magnification bias mentioned
earlier. The bias cannot be very strongly present in the  sample because
the lens systems with excessive magnifications are excluded
from the analysis.

Because the CSL analysis method is based on the flux ratios of the
images, the extinction effects from intergalactic matter
that affect both images in a similar way, can not change the results.
In principle, considerable amount of differential extinction present at
the lens object can corrupt the CSL value. However, the
lensed images are very often well outside the visible
lens object, which makes it improbable. 

At this point we can conclude that the same reason that makes
double image lenses unattractive targets for the substructure or
the micro-lensing studies gives them ideal properties as a probe for the
macro-lens profile. They are resilient against perturbations.

The main results from our study are the two halo profile populations.
We get cusp slope value $\alpha_{\mbox{\tiny H1}}=-1.95\pm 0.02$ and
normalization $n_{\mbox{\tiny H1}}=2.5\pm 0.7$
for population H1 and for population H2 correspondingly
$\alpha_{\mbox{\tiny H2}}=-1.49\pm 0.09$, $n_{\mbox{\tiny H2}}=0.5\pm 0.3$.
The normalization values indicate that 83 per cent of the lens objects
belong to population H1 and 17 per cent to population H2 with
4 per cent error margins  
in both values. This suggests that, on average, one sixth of the
lens objects in our sample belongs to the second
population with shallower cusp slope.
We performed a Monte Carlo study in order to find out the rejection
probability for the second population hypothesis. The result suggests
that the probability for the second population signature being noise
induced is at order of $\sim 1$ per cent.

As our Monte Carlo testing in section \ref{probs}
pointed out, the profiles with
shallower cusp slopes are more easily perturbed by error sources.
These deviations have tendency to make the acquired cusp slope
steeper than it really is. Although this is partly
reflected by rather wide
error range in the population H2 values, our referred population H2
results should be interpreted with caution. The cusp slope value
$\alpha_{\mbox{\tiny H2}}$ can be alleged as an estimate for the lower limit of
the population H2 profiles.

The foremost alternative explanation for the H2 population
is that they
are actually unresolved triple image lenses. However,
this scenario has several weaknesses: the fused pair of images 
should possess a strong autocorrelation in brightness variations that
is equal to the time delay between the images.
These unresolved lenses would increase the total
number of the triple image lenses and their corresponding statistical
frequency to such level that would contradict the acquired cusp slope
for the main population H1. If the H2 lenses were unresolved triple
image lenses in H1, the share of triple image lenses would be
11 out of 100, and the main population should
have profile $\alpha\approx -1.89$ according to the equation (\ref{tripleprob}).
This is well off from the determined profile value
for the main population.
Furthermore, although the H2 population of halos is smaller, they produce
distribution that is consistent with the model. This should not be the
case if they were triple image lenses -- their distribution should be
more random. 

According to the conventional wisdom acquired from the cosmological N-body
simulations, the universal cusp slope for the dark matter haloes is around
$\alpha\sim -1.0 ... -1.5$, depending on the author.  However,
observations on rotation curves of normal and LSB galaxies  suggest
even higher values, exceeding $\alpha=-1.0$. Our results contradict
most of these results. The lens density profiles has also
been studied by \cite{gava2007}. They used a power law profile with
joint weak and strong lensing modelling,
and confirmed the SIS lens population (H1),
but did not observe the shallower profile population H2.
The weak lensing does not have very good resolution, and the pure
power law profiles do not account the projected halo component outside
the scaling radius $r_s$ realistically. Additionally, they
study average properties of 22 lens halos. According to our results,
three or four lenses in their sample should have shallower profiles.
The second population
signature can easily be lost into the residual noise when
considering average properties of their lensing sample. 

Interestingly, \cite{gust2006} and \cite{roma2008} have
performed cosmological N-body simulations
with baryonic matter component. They have found similar populations for
the halo profiles as we do in our lens study. In the first population,
the halo profile is made steeper by the rich baryonic content, down to the value
$\alpha\sim -1.9$, which is close to the isothermal value.
They also discovered a second population of haloes
that were poor in baryonic content, with cusp value $\alpha\sim -1.3$.
It would seem, that although baryonic matter density is much lower
than density of dark matter, it can still have non-negligible
effects on the evolution of dark matter.

Our numerical values are also supported by the additional results
discussed in appendix \ref{cusplimderiv} and
the section \ref{tripleimage}.
Flat cores $\alpha>-1$ are very poor in
strong lensing, and shallow profiles $\alpha\gtrsim -1.5$
should produce excessive number of triple image lenses.

The CSL concept has clear advantages. The method uses more
  realistic lensing model inspired by N-body simulations.
The only required data is
 sampled double image lens photometry. The method can
 separate intrinsic properties of the lensing halos from the
 extrinsic lensing conditions. Thus no information on the lens and the source
 redshifts, the lens masses  or concentrations, cosmology, or
 theoretical distribution functions characterizing the lens and source
 populations is needed. This removes many uncertainties that are
 usually present in the statistical lensing studies.

The method has also disadvantages.
When general properties of dark matter haloes are considered,
there is a strong selection bias in our study. Our results concern
only those haloes that are able to do strong lensing, i.e. haloes
with cusp slopes steeper than $\alpha\sim -1$.
Generally, the lens objects are massive elliptical or early type 
galaxies residing at the core of a small galaxy cluster which would
incorporate the dark matter halo of the whole cluster into the lensing
potential. These kinds of objects are very different from the late type
spiral galaxies or LSB-galaxies, which usually cannot perform strong lensing
at all. Additionally, there are relatively few
observed double image lenses at the moment,
and the CSL method requires a hand-picked sample
of relatively unperturbed lens systems exhibiting axial symmetry.

New lens systems are found all the time. Certain proportion of these
are double image lenses. Larger data-set produces better
results, making the CSL analysis more accurate in the future.

\section*{Acknowledgments}

The author would like to acknowledge stimulating discussions with
 Pekka Hein\"am\"aki, Pasi Nurmi and Pertti Rautiainen on the key
topics in this paper. The author thanks also Marja Annala for her
support and help throughout the development of this work. Jussi Alarauhio
and Kari Rummukainen are thanked for carefully reading and commenting
an early version of this paper. The referee is thanked for the
constructive comments on this manuscript. The work of the author
has been supported in part through graduate student funding
of the University of Oulu. We are grateful for the funding provided by
the Magnus Ehrnrooth foundation. This research has made use
of the NASA Astrophysics Data System.

\appendix

\section{Cusp slope limit for strong lensing}
\label{cusplimderiv}

Our previously published lensing theory in \cite{muma2006}
approximates the surface density $\kappa$ produced by the GNFW mass profile
with
\begin{equation}
\kappa=ax^{\alpha+1}+b,
\end{equation}
where constants $a$ and $b$ depend only on the cusp slope value
$\alpha$ of the profile. The radial coordinate $x$ on the lens plane and
similar coordinate $y$ on the source plane are defined
as in \cite{schn1992}.
Corresponding un-normalized lens
equation is ($\alpha\ne-1$)
\begin{equation}
y=x-\frac{\mu_0}{x}\left(\frac{a|x|^{\alpha+3}}{(\alpha+3)}
+\frac b2|x|^2\right).
\end{equation}
Here $\mu_0=4\rho_0 r_s/\Sigma_{\tiny\mbox{cr}}$ is the normalized
projected density. This lens equation has its Einstein radius at $y=0$, i.e.
\begin{equation}\label{unnein}
x=x_{\mbox{\tiny e}}=\left[\frac{\alpha+3}{\mu_0a}\left(1-\frac{\mu_0b}{2}\right)\right]^{1/(\alpha+1)},
\end{equation}
where $\mu_0=4\rho_0r_{\mbox{\tiny s}}/\Sigma_{\mbox{\tiny cr}}$. Here the scale radius is
\begin{equation}
r_{\mbox{\tiny s}}=\frac{1}{c_1}\left\{\frac{3M}{800\pi\rho_{\mbox{\tiny cr}}}\right\}^{1/3},
\end{equation}
where $M$ is the total mass of the lens and $c_1$ the profile
concentration. The central density is
\begin{equation}
\rho_0=\frac{200c_1^3\rho_{\mbox{\tiny cr}}}{f(c_1)},
\end{equation}
where
\begin{equation}\label{funfun}
f(c_1)=\int\limits_0^{c_1}\mbox{d}x\; x^{\alpha+2}(1+x)^{-(\alpha+3)}.
\end{equation}
In the previous equations, the critical density $\rho_{\mbox{\tiny cr}}$ is defined as
\begin{equation}\label{pim1}
\rho_{\mbox{\tiny cr}}=\frac{3H_0^2}{8\pi G}\left[\Omega_m(1+z_{\mbox{\tiny L}})^3+\Omega_{\mbox{\tiny R}}(1+z_{\mbox{\tiny L}})^2+\Omega_\Lambda\right].
\end{equation}
and the critical surface density is
\begin{equation}\label{pim2}
\Sigma_{\mbox{\tiny cr}}=\frac{c^2 D_{\mbox{\tiny S}}}{4\pi G D_{\mbox{\tiny L}} D_{\mbox{\tiny LS}}}.
\end{equation}
In equations (\ref{pim1}) and (\ref{pim2}) the Hubble constant is
$H_0$, the gravitational constant is $G$ and the speed of light is
$c$. Subscripts $L$ and $S$ refer to the lens and the source, and symbol $D$
to corresponding angular cosmological distance and $z$ to  redshift.
The total energy density of the universe is parametrized with $\Omega_m$,
$\Omega_{\mbox{\tiny R}}$ and $\Omega_\Lambda$ (matter, curvature and cosmological constant).

At the limit where the lens cannot produce multiple images, the Einstein radius
is reduced to zero and becomes imaginary.
Thus, a condition for strong
lensing can be acquired from the equation (\ref{unnein}) as
\begin{equation}\label{strongcond1}
\mu_0<\frac{2}{b}.
\end{equation}
This equation can be solved numerically with the definitions presented
above, but it can also be approximated with
\begin{equation}
b\approx(\alpha+1)^{-1}(\alpha+2)^{-1}
\end{equation}
and
\begin{equation}
f(c_1)=f(c_1;\alpha)\approx
f(c_1;-1)=\log(1+c_1)-\frac{c_1}{1+c_1}
\end{equation}
with good accuracy when $\alpha>-1$. When these approximations are
inserted into the condition (\ref{strongcond1}), we can solve the
maximum cusp slope for the strong lensing as
\begin{equation}
\alpha=\alpha_{\mbox{\tiny sl}}=-\frac{3}{2}+\frac 12\sqrt{
1+\frac{1600 r_{\mbox{\tiny s}}\rho_{\mbox{\tiny cr}}c_1^3(1+c_1)}{3\Sigma_{\mbox{\tiny cr}}[(1+c_1)\log(1+c_1)+c_1]}
}.
\end{equation}

\label{lastpage}

\end{document}